\documentclass[journal,twoside]{IEEEtran}

\usepackage{cite,amsmath,amssymb}
\usepackage{subfigure}
\usepackage{citesort}
\usepackage{fancyhdr}

\usepackage{ifpdf}

\ifpdf
    \usepackage[pdftex]{graphicx}
    \usepackage[pdftex]{hyperref}
\else
    \usepackage{graphicx}
\fi

\ifpdf \hypersetup{    backref=section,
                colorlinks=true,
                plainpages=false,
                breaklinks=true,
                colorlinks=true,
                urlcolor=blue,
                citecolor=blue,
                linkcolor=blue,
                bookmarks=true,
                pdfpagemode=none,
                pdfstartview=FitH}
\fi

\newtheorem{theorem}{Theorem}

\newtheorem{lemma}{Lemma}

\newtheorem{corollary}{Corollary}
\newenvironment{corollarybox} {\begin{corollary}}{\hfill \interlinepenalty500 $\Box$\end{corollary}}

\newlength{\oldparindent}
\newlength{\oldparskip}
\newcommand{\papertitle}{On the Mutual Information Distribution of OFDM-Based Spatial Multiplexing: Exact Variance and Outage Approximation}

\newlength{\figwidth}
\newif\ifpdf
\ifx\pdfoutput\undefined
    \pdffalse 
\else
    \pdfoutput=1 
    \pdftrue
\fi

\ifpdf
    \usepackage[pdftex]{graphicx}
    \usepackage[pdftex]{hyperref}
\else
    \usepackage{graphicx}
\fi

\ifpdf \hypersetup{    backref=section,
                colorlinks=true,
                plainpages=false,
                breaklinks=true,
                colorlinks=true,
                urlcolor=blue,
                citecolor=blue,
                linkcolor=blue,
                bookmarks=true,
                pdfpagemode=none,
                pdfstartview=FitH}
\fi

\newcommand{\defeq}{\stackrel{\Delta}{=}}

\newcommand{\bH}{\mathbf{H}}

\begin{document}

\title{\papertitle}
\author{Matthew R.\ McKay,~\IEEEmembership{Member,~IEEE}, Peter J.\ Smith,~\IEEEmembership{Senior Member,~IEEE}, Himal A.\ Suraweera~\IEEEmembership{Member,~IEEE} \; and \;  Iain
B.~Collings,~\IEEEmembership{Senior Member,~IEEE}
\thanks{Manuscript received May 12, 2006; revised July 11, 2007.
This paper was presented in part at IEEE ICC, Glasgow, UK, 2007.}%
\thanks{M.\ R.\ McKay is with the Department of Electronic and Computer Engineering, Hong Kong University of Science and Technology, Clear Water Bay, Kowloon, Hong Kong (e-mail: eemckay@ust.hk).}
\thanks{P.\ J.\ Smith is with the Department of Electrical and Computer Engineering,
University of Canterbury, Private Bag 4800, Christchurch, New Zealand (e-mail: p.smith@elec.canterbury.ac.nz).}
\thanks{H.\ A.\ Suraweera was with the Department of Electrical and Computer Engineering, Monash University, Melbourne, Victoria 3800, Australia.  He is now with the Center for Telecommunications and Microelectronics (T$\mu$E), Victoria University, PO Box 14428, Melbourne City MC, Victoria 8001, Australia (email: himal.suraweera@vu.edu.au).}
\thanks{I.\ B.\ Collings is with the ICT Centre, CSIRO, NSW, Australia. (e-mail: Iain.Collings@csiro.au).}
} \markboth{IEEE Transactions on Information
Theory (To Appear)}{\papertitle}  

\maketitle

%
%
%

\begin{abstract}
This paper considers the distribution of the mutual information of
frequency-selective spatially-uncorrelated Rayleigh fading MIMO
channels.  Results are presented for OFDM-based spatial
multiplexing.  New exact closed-form expressions are derived for the
variance of the mutual information.  In contrast to previous
results, our new expressions apply for systems with both arbitrary
numbers of antennas and arbitrary-length channels. Simplified
expressions are also presented for high and low SNR regimes. The
analytical variance results are used to provide accurate analytical
approximations for the distribution of the mutual information and
the outage capacity.

\end{abstract}


\section{Introduction} \label{sec:Intro}

Multiple-input multiple-output (MIMO) antenna technology has emerged
as an effective technique for significantly improving the capacity
of wireless communication systems. A great deal of work has been
done on analyzing the MIMO capacity in various flat-fading channel
scenarios, since the pioneering work of \cite{teletar99} and
\cite{foschini96}. In particular, the mean (ergodic) capacity has
now been comprehensively investigated (e.g.\ see
\cite{grant02,chiani03,smith03,shin03,lozano03,moustakas_03,kiessling04,simon_04_submit,alfano04_2,lozano05_jnl,mckay_jnl05,tulino05_IT,kang04,mck06_let,mckay06_ISIT,tulino04_jnl}
and references therein).

In addition, the outage capacity has also been investigated for
flat-fading channels. This is an important capacity measure for
systems with stringent delay constraints, and also provides
information about the system diversity \cite{bolcskei02}. With the
exception of the exact two/three antenna results presented in
\cite{smith03_letter,smith04},
outage capacity analysis has typically involved approximating the
distribution of the mutual information, since exact closed-form
solutions are not forthcoming. It has been shown that the Gaussian
distribution provides a good approximation in many cases
\cite{smith02,smith03,moustakas_03,wang04_MI,mckay_jnl05}.

\begin{table}
\caption{MIMO-OFDM presence in standards}
\label{table_1}
\begin{center}
\begin{tabular}{|c|c|}
\hline
Standard & Technology\\
\hline
\hline
WLAN IEEE 802.11n & OFDM \\
\hline
WiMAX IEEE 802.16-2004& OFDM/OFDMA\\
\hline
 WiMAX IEEE 802.16e&  OFDMA   \\
\hline
MBWA IEEE 802.20 &  OFDM         \\
\hline
WRAN IEEE 802.22 &    OFDM       \\
\hline
3GPP Release 8 &     OFDMA               \\
\hline
\end{tabular}
\end{center}
\end{table}


In this paper, we consider frequency-selective MIMO channels, which
are applicable for many current high data-rate wireless systems.  We focus on MIMO
orthogonal frequency-division multiplexing (OFDM) systems, since they form the underlying technology for a many emerging MIMO standards, as summarized in Table \ref{table_1}, and consider spatial multiplexing transmission. Despite their key practical significance however, for these systems (and indeed frequency-selective MIMO channels in general) there are relatively few analytic MIMO capacity results.  The ergodic capacity (average mutual information) was considered in
\cite{bolcskei02,wang02_ofdm,oyman03,liu03_IT} and \cite{pillutla04,mckay_jnl06}, assuming Rayleigh and Rician channels
respectively, and was found to be easily obtained by summing the equivalent flat-fading ergodic MIMO capacity of each individual OFDM
subcarrier. In contrast, the outage capacity does not decompose in this way.


Calculating the outage capacity for frequency-selective channels is
difficult due to the non-negligible correlations between subcarrier
channel matrices. As such, the investigation of outage capacity has
usually been performed using simulation studies
\cite{bolcskei02,kafle04,suraweera05}. It appears that the only
current analytical outage capacity results for frequency-selective
channels are presented in \cite{Clark07}, \cite{barriac04} and
\cite{moustakas06}, all of which derive a Gaussian approximation for the mutual information distribution.  The results in \cite{Clark07} however, are based on deriving exact expressions for the mutual information variance of single-input single-output (SISO) channels only;  whereas the results in \cite{barriac04} and \cite{moustakas06} are based on approximating the mutual information variance using asymptotic methods. Specifically, \cite{barriac04} considers multiple-input single-output (MISO) channels with asymptotically large channel lengths, whereas \cite{moustakas06} considers MIMO channels with infinite numbers of transmit and receive antennas.
We note also that for the extreme frequency-selective fading case, ie.\ where the MIMO subcarrier matrices are independent across frequency, the variance of the mutual information could be easily calculated by adapting known MIMO flat-fading variance results given, for example, in \cite{moustakas_03} and \cite{kang04}. For many practical systems however, the subcarrier channels are typically highly correlated across frequency, and this approach cannot be applied.

In this paper, we consider MIMO OFDM-based spatial multiplexing systems
with finite numbers of antennas, and operating over spatially-uncorrelated Rayleigh fading channels with
finite delay spreads. We first derive new exact closed-form
expressions for the mutual information variance. We also give
explicit reduced formulas for the specific cases of multiple-input
single-output (MISO), single-input multiple-output (SIMO), and
single-input single-output (SISO) systems. Moreover, simplified
closed-form expressions are derived for the variance in the high and
low signal-to-noise ratio (SNR) regimes.

Based on the new analytic variance results (along with known
analytic mean results), we then present new approximations to the
mutual information distribution of OFDM-based spatial multiplexing
systems. In particular, we present a new closed-form Gaussian
approximation, which is shown to be extremely accurate for many
different system and channel scenarios.  In the low SNR regime, we
also present a new analytic Gamma approximation, which we show to be
more accurate than the Gaussian approximation in this case.

Finally, we use the analytic Gaussian approximation to estimate the
outage capacity.  We find that the approximation is very accurate,
and show that for outage levels of practical interest, the outage
capacity depends heavily on the delay spread of the channel.

The paper is organized as follows. In Section II we describe the
frequency-selective MIMO channel model, the OFDM-based spatial
multiplexing signal model, and the associated mutual information. In
Section III, we present the main analytical contributions of the
paper, namely, analytical expressions for the variance of the mutual
information.  The proofs are relegated to the appendices. In Section
IV, we approximate the distribution of the mutual information, and
investigate the outage capacity.

The following notation is used throughout this paper. Matrices are
represented with uppercase boldface, and vectors with lowercase
boldface. The superscripts $(\cdot)^T, (\cdot)^*$, and
$(\cdot)^\dagger$ indicate matrix transpose, complex conjugate, and
complex conjugate transpose respectively. The matrix
$\mathbf{I}_{p}$ denotes a $p \times p$ identity matrix. We use
$\det\left( \cdot \right)$ and ${\rm tr}(\cdot)$ to represent the
matrix determinant and trace operations respectively. The operator
$E\left[\cdot\right]$ denotes expectation, and ${\rm Var}(\cdot)$
denotes variance.  The real Gaussian distribution with zero-mean and
unit-variance is denoted $\mathcal{N}(0,1)$, the corresponding
complex circularly symmetric Gaussian distribution is denoted
$\mathcal{CN}(0,1)$, and the chi-square distribution with $r$
degrees of freedom is denoted $\chi^2_r$.

\section{OFDM-Based Spatial Multiplexing Systems}

\subsection{Channel and Signal Model} \label{sec:ChanModel}

We consider a single-user OFDM-based spatial multiplexing system
employing $N_t$ transmit antennas, $N_r$ receive antennas, and $N$
subcarriers. The channel is assumed to be frequency-selective and is
modeled as a length-$L$ finite impulse-response (FIR) filter (as in
\cite{bolcskei02,oyman03}), for which the discrete-time input-output
relation is given by \cite{oyman03}
\begin{align} \label{eq:ioModel}
\mathbf{y}[q] = \sum_{p = 0}^{L-1} \sigma_p \mathbf{H}[p]
\mathbf{x}[q-p] + \mathbf{n}[q]
\end{align}
where $\mathbf{x}[q] \in \mathcal{C}^{N_t \times 1}$ is the signal
vector transmitted at sample index $q$,
 $\mathbf{y}[q] \in \mathcal{C}^{N_r \times 1}$ is the corresponding received signal
vector, and $\mathbf{n}[q] \in \mathcal{C}^{N_r \times 1}$ is the
noise vector containing independent elements $\sim
\mathcal{CN}(0,1)$. Also, $\sigma_p$, for $p = 0, \ldots, L-1$,
represents the channel power delay profile, and is normalized
according to
\begin{align}
\sum_{p = 0}^{L-1} \sigma_p^2 = 1 \; .
\end{align}

The $N_r \times N_t$ random matrices $\mathbf{H}[p]$, for $p = 0,
\ldots, L-1$, represent the MIMO channel impulse response. These
matrices are assumed to be mutually uncorrelated, and are assumed to
be known perfectly at the receiver but are unknown at the
transmitter.. The channel is assumed to be quasi-static, remaining
constant for the duration of a codeword, but changing independently
from codeword to codeword. Throughout the paper, we assume that the
channel elements exhibit spatially-uncorrelated Rayleigh
fading\footnote{Note that a number of recent investigations have
studied the impact of spatial correlation on MIMO capacity (see,
eg.\ \cite{moustakas_03,kiessling04,lozano05_jnl,mckay_jnl05}).  We
do not follow this line of work here however, since our primary
focus is to study the impact of frequency-selective fading on
capacity, in which case the effect of correlation is observed across
frequency.}, in which case each $\mathbf{H}[p]$ contains independent
elements $\sim \mathcal{CN}(0,1)$.

At the transmitter, the time-domain input sequence $\mathbf{x}[q]$
is generated as $N_t$ parallel OFDM symbols.
The symbols for each antenna are OFDM modulated using an $N$-point inverse fast-Fourier transform (IFFT)
prior to transmission. At the receiver, OFDM demodulation is
performed at each receive antenna using an $N$-point FFT. A key
advantage of OFDM-based spatial multiplexing is that equalization is
simple, since the frequency-selective MIMO channel is transformed
into $N$ orthogonal flat-fading MIMO subchannels via the IFFT/FFT
processing.

To maintain orthogonality in the presence of intersymbol
interference caused by multipath, OFDM systems typically employ a
cyclic prefix extension. Assuming that the cyclic prefix is longer
than the delay spread of the channel, we can write the equivalent
frequency domain input-output model for OFDM-based spatial
multiplexing as follows
\begin{align} \label{eq:ioModel_freq}
\mathbf{r}_k = \mathbf{H}_k \mathbf{a}_k + \mathbf{n}_k, \; \; \; k
= 0, \ldots, N-1
\end{align}
where $\mathbf{a}_k$ is the transmitted vector for the $k$th
subcarrier, assumed to be i.i.d.\ Gaussian with covariance matrix $E \bigl[ \mathbf{a}_k \mathbf{a}_k^\dagger
\bigr] = \frac{\gamma}{N_t} \mathbf{I}_{N_t} $, $\mathbf{r}_k$ is
the received vector for the $k$th subcarrier, and $\mathbf{n}_k$ is
the corresponding complex AWGN vector satisfying
$E \bigl[ \mathbf{n}_k \mathbf{n}_\ell^\dagger \bigr] =
\mathbf{I}_{N_r} \delta [k - \ell]$,
where $\delta[\cdot]$ is the Kronecker-delta function. Also,
$\mathbf{H}_k$ is the $k$th subcarrier channel matrix given by
\begin{equation} \label{eq:OFDM_channel_mat}
\bH_k = \sum_{p = 0}^{L-1} \sigma_p \, \bH [p ] \exp \left( -j
2\pi \frac{k}{N} p \right)
\end{equation}
containing independent entries $(\mathbf{H}_k)_{i,j} \sim
\mathcal{CN}(0,1)$.  Note that due to the finite-length impulse
response, correlation exists between different subcarrier channel
matrices. Using (\ref{eq:OFDM_channel_mat}), the correlation
coefficients between the channel elements on two arbitrary
subcarriers $k$ and $\ell$ is easily derived as follows (see also
\cite{li98})
\begin{align} \label{eq:freqCorrCoef}
\rho_{k - \ell} &= E \left[ (\mathbf{H}_k)_{i,j}
(\mathbf{H}_\ell)_{i',j'}^* \right] \nonumber \\
&=  \sum_{p = 0}^{L-1} \sigma_p^2
\; e^{-j 2 \pi (k-\ell) p / N } \delta[i-i'] \delta[j-j']
\end{align}
for all $i,j,i',j'$. As expected, these frequency correlation
coefficients depend only on the difference between subcarriers
(i.e.\ $k - \ell$), and not on the subcarriers themselves.

Note that with the above model, the SNR per receive antenna per
subcarrier (henceforth referred to as `the SNR') is given by
$\gamma$.

\subsection{Mutual Information}

The focus of this paper is on the statistics of the mutual
information of OFDM-based spatial multiplexing systems.  It is now
well-known that the instantaneous mutual information in b/s/Hz for a
given channel realization is given by \cite{bolcskei02}
\begin{align} \label{eq:OFDMMi}
\mathcal{I}_{\rm ofdm} = \frac{1}{N} \sum_{k=0}^{N-1} \mathcal{I}_k
\end{align}
where $\mathcal{I}_k$ is the instantaneous mutual information for
the $k$th OFDM subcarrier, given by
\begin{align} \label{eq:logDet}
\mathcal{I}_k = \log_2 \det \left( \mathbf{I}_{N_r} +
\frac{\gamma}{N_t} \mathbf{H}_k \mathbf{H}_k^\dagger \right) \; .
\end{align}
Note that the loss in mutual information due to the cyclic prefix
has been neglected in (\ref{eq:logDet}). The mean (ergodic) mutual
information is given by
\begin{align} \label{eq:meanMI}
E \left[ \mathcal{I}_{\rm ofdm} \right] = \frac{1}{N}
\sum_{k=0}^{N-1} E \left[ \mathcal{I}_k \right] \; .
\end{align}
It is obvious that (\ref{eq:meanMI}) is equivalent to the ergodic
mutual information of a flat-faded channel, for which case closed-form expressions are now available \cite{shin03,Dohler03,kang04}.

\section{Variance of the Mutual Information}

In this section we derive new closed-form expressions for the
variance of the mutual information of OFDM-based spatial
multiplexing.  Our results are exact, and apply for arbitrary finite
system and channel parameters. We also present simplified
expressions for the variance in the high and low SNR regimes, and
give explicit reduced variance expressions for the cases of MISO,
SIMO, and SISO systems.  These results will be subsequently used in
Section \ref{sec:outageApprox} for providing accurate approximations
to the mutual information distribution, and to the outage capacity.

\subsection{Exact Analysis at All SNRs}

The following theorem presents an exact expression for the variance
of the mutual information of MIMO-OFDM systems.
\begin{theorem} \label{th:exactVarAllSNR}
The variance of the mutual information of MIMO-OFDM systems is given
by
\begin{align} \label{eq:exactVarianceFinal}
{\rm Var}(\mathcal{I}_{\rm ofdm}) &= \frac{ (\log_2 (e))^2
}{\Gamma_m(n) \Gamma_m(m)} \biggl( \frac{2}{N^2} \sum_{d=1}^{N-1}
(N-d) \varphi(\rho_d) \nonumber \\
& + \sum_{r = 1}^m \sum_{s = 1}^m \frac{ \det
\left( \mathbf{B}_{r,s} \right)}{N} - \frac{\bigl( \sum_{r=1}^m \det
\left( \mathbf{A}_r \right) \bigr)^2}{\Gamma_m(n) \Gamma_m(m)}
 \biggr)
\end{align}
where $m = \min(N_r, N_t), n = \max(N_r, N_t)$, $\Gamma_m(\cdot)$ is
the complex multivariate gamma function defined as
\begin{align} \label{eq:mvgamDefn}
\Gamma_m \left( n \right) = \prod\limits_{i = 1}^m {\Gamma \left( {n
- i + 1} \right)}
\end{align}
and
\begin{align} \label{eq:PhiFcn_defn}
\varphi(\rho_d) = \left\{
\begin{array}{ll}
\frac{\bigl( \sum_{r=1}^m \det \left(
\mathbf{A}_r \right) \bigr)^2}{\Gamma_m(n) \Gamma_m(m)} &, \, | \rho_d| = 0 \\
\sum_{r = 1}^m \sum_{s = 1}^m e^{2 N_t / \gamma} \det \left(
\mathbf{C}_{r,s} (\rho_d) \right)  &, \,   0 < |\rho_d| < 1 \\
\sum_{r = 1}^m \sum_{s = 1}^m  \det \left( \mathbf{B}_{r,s} \right)
  &, \,  |\rho_d| = 1
\end{array}
\right.
\end{align}
The matrix $\mathbf{A}_r$ is $m \times m$, with $(i,j)^{\rm th}$
element
\begin{align} \label{eq:Ar_defn}
\left(\mathbf{A}_{r} \right)_{i,j} = \left\{
\begin{array}{ll}
b!  & \text{for} \; \; j \neq r \\
b! e^{N_t/\gamma} g_1(b+1) & \text{for} \; \;  j = r \\
\end{array}
\right. \; .
\end{align}
The matrices $\mathbf{B}_{r,s}$ and
$\mathbf{C}_{r,s} (\cdot)$ are $m \times m$ with $(i,j)^{\rm th}$ elements given by (\ref{eq:Brs_defn}) and (\ref{eq:Ckell_defn}) respectively (at the top of the next page).
\newcounter{mytempeqncnt}
\begin{figure*}[!t]
\setcounter{mytempeqncnt}{\value{equation}}
\setcounter{equation}{12}
\begin{align} \label{eq:Brs_defn}
&\left(\mathbf{B}_{r,s} \right)_{i,j} = \left\{
\begin{array}{ll}
b!  &, \; \; {\rm for} \; \;  j \neq r \; {\rm and} \; j \neq s \\
b! e^{N_t/\gamma} g_1(b+1)  &, \; \; {\rm for} \; \;  j = r \; {\rm or} \; j = s, \; {\rm and} \; r \neq s \\
2 \left(N_t/\gamma\right)^{b+1} e^{N_t/\gamma} \sum_{t=0}^b \binom{b}{t} (-1)^t \\ \hspace*{0.1cm} \times {\rm G}_{3,4}^{4,0}
 \left( N_t/\gamma \bigr|_{0, t-b-1, t-b-1, t-b-1}^{t-b, t-b, t-b} \right) & \; {\rm for} \; \;, \;  j = r = s \\
\end{array}
\right.
\end{align}
\setcounter{equation}{\value{mytempeqncnt}}
\hrulefill
\vspace*{4pt}
\end{figure*}
\begin{figure*}[!t]
\setcounter{mytempeqncnt}{\value{equation}}
\setcounter{equation}{13}
\begin{align} \label{eq:Ckell_defn}
\left(\mathbf{C}_{r, s}(\rho_d) \right)_{i,j} = \left\{
\begin{array}{ll}
\eta_{i,j} (1, \rho_d)  & \text{for} \; \; i \neq r, \; j \neq s \\
\eta_{i,j} (g_1(z), \rho_d) & \text{for} \; \;  i = r, \; j \neq s \\
|\rho_d|^{2(i-j)} \, \eta_{j,i} (g_1(z), \rho_d) & \text{for} \; \;  i \neq r, \; j = s \\
\frac{(1-|\rho_d|^2)^{z}}{|\rho_d|^{2 (j-1)}} e^{ \frac{ 2 N_t
|\rho_d|^2 }{ \gamma (1 - |\rho_d|^2)}} \sum_{t=0}^\infty
\frac{ |\rho_d|^{2 t} \Gamma(u) \Gamma(v) g_2(u) g_2(v) }{t! (\tau + t)!}  & \text{for} \; \; i = r, \; j = s \\
\end{array}
\right. \;
\end{align}
\setcounter{equation}{\value{mytempeqncnt}}
\hrulefill
\vspace*{4pt}
\end{figure*}
\addtocounter{equation}{2}
Also, $b = n + m - i - j$, $\tau = n - m$, $z = \tau +
i + j -1$, $u = \tau + i + t$, $v = \tau + j + t$, and ${\rm
G}_{3,4}^{4,0}(\cdot)$ is the Meijer-G function (see \cite[eq.\
(9.301)]{gradshteyn65} for definition),
\begin{align} \label{eq:suxDefn}
&g_1(z) =  \sum_{h=1}^z {\rm E}_h \left( \frac{N_t}{\gamma} \right) \\
&g_2(z) =  \sum_{h=1}^z {\rm E}_h \left( \frac{N_t}{\gamma \, (1 -
|\rho_d|^2)} \right)
\end{align}
where ${\rm E}_h(\cdot)$ is the Exponential Integral (see \cite[eq.
(5.1.12)]{abramowitz65} for definition).  The function $\eta_{i,j}(\cdot, \cdot)$ is
defined as
\begin{align} \label{eq:EtaDefn}
\eta_{i,j}(f(z), \rho_d) &= \Gamma(\tau+j) \sum_{t=0}^{j-1}
\binom{j-1}{t} \left(\frac{ 1 - |\rho_d|^2}{ |\rho_d|^2}\right)^t \nonumber \\
& \times (\tau + j - t)_{i-1} \, f(z-t)
\end{align}
for an arbitrary input function $f$, and $(\cdot)_r$ is the
Pochammer symbol
\begin{align}
(a)_r = a \cdot (a + 1) \cdot \ldots \cdot (a+r-1) =
\frac{\Gamma(a+r)}{\Gamma(a)} ; \; \; \; \; (a)_0 = 1 \; .
\end{align}
\end{theorem}

\emph{Proof:} See Appendix \ref{app:ProofExactVar}. \hfill
\interlinepenalty500 $\Box$

\begin{figure*}[!t]
\setcounter{mytempeqncnt}{\value{equation}}
\setcounter{equation}{20}
\begin{align}
\varphi(\rho_d) = \left\{
\begin{array}{ll}
\Gamma(n) e^{2 N_t / \gamma} ( g_1(n) )^2  & \; \; \text{for} \; \; | \rho_d| = 0 \\
(1-|\rho_d|^2)^n \, e^{\frac{2 N_t}{\gamma (1 - |\rho_d|^2)}}
\sum_{t=0}^\infty \frac{ |\rho_d|^{2 t} \Gamma(n+t) (
g_2(n+t))^2}{t!}  & \; \; \text{for} \; \;   0 < |\rho_d| < 1 \\
2 e^{N_t/\gamma} \left( \frac{N_t}{\gamma}\right)^n \,
\sum_{t=0}^{n-1} \binom{n-1}{t} (-1)^{n-1-t} {\rm G}_{3,4}^{4,0}
\left( N_t/\gamma \bigr|^{-t, -t, -t}_{0, -t-1, -t-1, -t-1} \right)
  & \; \; \text{for} \; \;  |\rho_d| = 1
\end{array}
\right.
\label{eq:varphiDefn}
\end{align}
\setcounter{equation}{\value{mytempeqncnt}}
\hrulefill
\vspace*{4pt}
\end{figure*}

Note that the exact variance expression in Theorem \ref{th:exactVarAllSNR} can be easily evaluated since it primarily involves simple polynomial and exponential terms, as well as standard functions such as exponential integrals and Meijer-$G$ functions, both of which are implemented as built-in procedures in various mathematical software packages such as Maple and Mathematica.
We also note that although Theorem
\ref{th:exactVarAllSNR} involves infinite series of exponential
integrals, its numerical evaluation can be made more efficient by
exploiting the following recurrence relations \cite[eqs.\ (5.1.7)
and (5.1.14)]{abramowitz65}
\begin{align}
& E_1 (z) = - \text{Ei} (-z) \nonumber \\
& E_{n+1} (z) = \frac{1}{n} \left( e^{-z} - z E_n (z) \right)
\end{align}
for $z > 0$.  As such, only a single exponential integral must be
explicitly evaluated when summing these series.
Moreover, it turns out that this infinite series converges quickly, and can generally be evaluated with less than $20$ terms.  Therefore the computational challenge associated with this series is very low.

The following corollary presents an exact variance expression for the
mutual information of SIMO and MISO OFDM systems (i.e.\ cases with
$m = 1, n > 1$). To the best of our knowledge, this result is also new.

\begin{corollarybox} \label{corr:exactVarSIMOAllSNR}
The variance of the mutual information of SIMO/MISO-OFDM systems is
given by
\begin{align} \label{eq:exactVarianceFinalMISO}
{\rm Var}(\mathcal{I}_{\rm ofdm}) &=  \frac{(\log_2
(e))^2}{\Gamma(n)} \biggl( \frac{2}{N^2} \sum_{d=1}^{N-1} (N-d)
\varphi(\rho_d) \nonumber \\
& \hspace*{-1.5cm} - \Gamma(n) e^{2 N_t / \gamma} ( g_1(n) )^2  + \frac{2 e^{N_t/\gamma}}{N} \left(
\frac{N_t}{\gamma}\right)^n  \nonumber \\
& \hspace*{-1.5cm} \times \sum_{t=0}^{n-1} \binom{n-1}{t}
(-1)^{n-1-t} {\rm G}_{3,4}^{4,0} \left( N_t/\gamma \bigr|^{-t, -t,
-t}_{0, -t-1, -t-1, -t-1} \right)
 \biggr)
\end{align}
where $\varphi(\rho_d)$ is defined in (\ref{eq:varphiDefn}) at the top of the page.

\addtocounter{equation}{1}

\end{corollarybox}

The following corollary presents an exact variance expression for the
mutual information of SISO OFDM systems (i.e. $m = 1, n = 1$).

\begin{corollarybox} \label{corr:exactVarSISOAllSNR}
The variance of the mutual information of SISO-OFDM systems is given
by
\begin{align} \label{eq:exactVarianceFinalSISO}
{\rm Var}(\mathcal{I}_{\rm ofdm}) &=  (\log_2 (e))^2 \biggl(
\frac{2}{N^2} \sum_{d=1}^{N-1} (N-d) \varphi(\rho_d) \nonumber \\
& \hspace*{-1cm} + \frac{2
e^{1/\gamma}}{\gamma N} {\rm G}_{3,4}^{4,0} \left( 1/\gamma
\bigr|^{\; 0, \; 0, \; 0}_{0, -1, -1, -1} \right) - e^{2 / \gamma} (
g_1(1) )^2
 \biggr)
\end{align}
where
\begin{align}
\varphi(\rho_d) = \left\{
\begin{array}{ll}
e^{2 / \gamma} ( g_1(1) )^2  & \text{for} \; \, | \rho_d| = 0 \\
(1-|\rho_d|^2) \, e^{\frac{2}{\gamma (1 - |\rho_d|^2)}} \\
\; \; \times \sum_{t=0}^\infty |\rho_d|^{2 t} ( g_2(1+t))^2  & \text{for} \; \,  0 < |\rho_d| < 1 \\
\frac{2 e^{1/\gamma}}{\gamma N} {\rm G}_{3,4}^{4,0} \left( 1/\gamma
\bigr|^{\; 0, \; 0, \; 0}_{0, -1, -1, -1} \right)
  & \text{for} \; \,  |\rho_d| = 1
\end{array}
\right.
\end{align}

\end{corollarybox}

\begin{figure*}[!t]
\setcounter{mytempeqncnt}{\value{equation}}
\setcounter{equation}{26}
\begin{align}
\left(\mathbf{\tilde{C}}_{r, s}(\rho_d) \right)_{i,j} &= h^2(\rho_d)
|\rho_d|^{2(i-j)} \, \eta_{j,i} (1, \rho_d) + |\rho_d|^{2(i'-j)}
\eta_{j',i'}
(\xi_{|\rho_d|^2} (z-1), \rho_d) + h(\rho_d) \bigl( \eta_{i,j} (H(z-1), \rho_d) \nonumber \\
& \hspace*{1cm} - \ln(1 - |\rho_d|^{2}) \eta_{i,j} (1, \rho_d) +
|\rho_d|^{2(i-j)} ( \eta_{j,i} (H(z-1), \rho_d) - \ln(1 -
|\rho_d|^{2})  \eta_{j,i} (1, \rho_d)  )  \bigr)
\label{eq:Ctilde_rs_Simple}
\end{align}
\setcounter{equation}{\value{mytempeqncnt}}
\hrulefill
\vspace*{4pt}
\end{figure*}

Very recently, an equivalent expression for the SISO-OFDM variance given in (\ref{eq:exactVarianceFinalSISO}) was presented in \cite{Clark07}\footnote{Note that this expression was not explicitly stated in \cite{Clark07}. It can however be trivially obtained by following the derivation of (\ref{eq:VarEqn}) and using \cite[Eqs.\ (12), (41), and (48)]{Clark07}.}.  In contrast to (\ref{eq:exactVarianceFinalSISO}) however, the equivalent result from \cite{Clark07} is not expressed in closed-form, and it requires the evaluation of infinite series of incomplete gamma functions.

\begin{figure} \centering
\includegraphics[width=\figwidth]{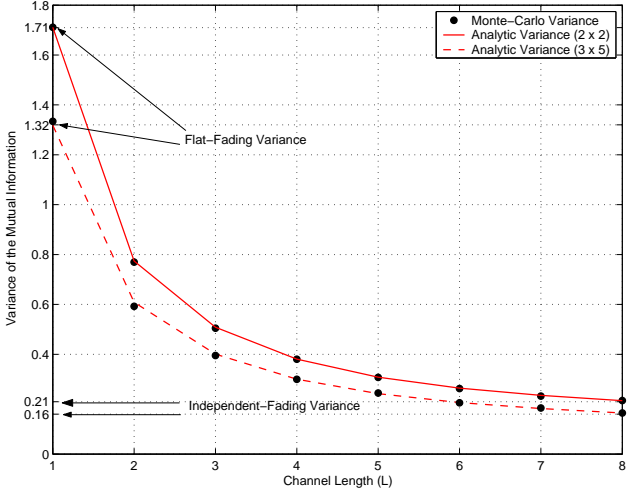}
\caption{Variance of the mutual information of MIMO-OFDM  for
different $N_t \times N_r$ antenna configurations, and different
channel lengths (uniform power delay profile). The ``Analytic
Variance'' curves are based on (\ref{eq:exactVarianceFinal}). $8$
subcarriers is considered, with SNR of $10$ dB.}
\label{fig:var_Diffant}
\end{figure}

In Fig.\ \ref{fig:var_Diffant} we compare the analytical variance
expression (\ref{eq:exactVarianceFinal}) with the variance obtained
via Monte-Carlo simulation.  Results are presented for two different
$N_t \times N_r$ antenna configurations as a function of the channel
length $L$.  A uniform power delay profile is assumed (i.e.\
$\sigma^2_p = 1/L$, for $p = 0, \ldots L-1$), $8$ subcarriers are
used (simply by way of example; similar results are obtained for
higher numbers of subcarriers), and the SNR is set to $10 {\rm dB}$.
In all cases we see a precise agreement between the simulated and
analytic curves. Moreover, the variance is seen to be largest for
the system with the least antennas, regardless of the channel
length. For both antenna configurations, we see that the variance
reduces with increasing $L$, and that this reduction is most
significant for small $L$. For example, by increasing the channel
length from $L=1$ (flat-fading) to $L=2$, the variance for both
antenna configurations is more than halved.

\begin{figure} \centering
\includegraphics[width=\figwidth]{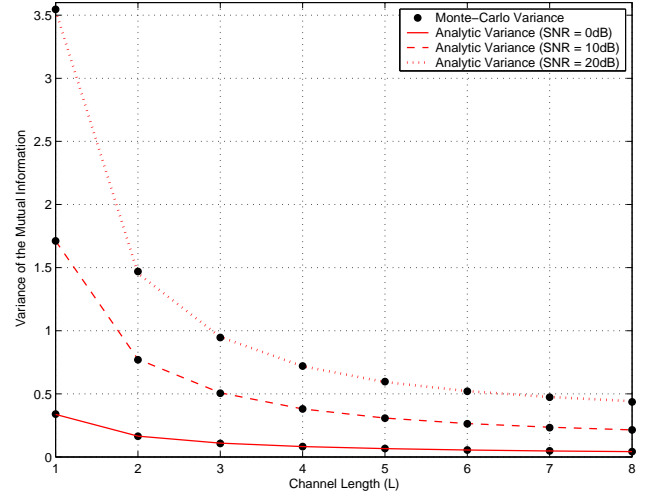}
\caption{Variance of the mutual information of MIMO-OFDM  for
different SNRs, and different channel lengths (uniform power delay
profile). The ``Analytic Variance'' curves are based on
(\ref{eq:exactVarianceFinal}). A $2 \times 2$ system is considered
with $8$ subcarriers.} \label{fig:var_Diffsnr}
\end{figure}

In Fig.\ \ref{fig:var_Diffsnr} we plot the analytical variance
expression (\ref{eq:exactVarianceFinal}) and Monte-Carlo simulation
results for different SNRs, as a function of $L$.  Again we see a
precise agreement between the analytical and simulated results. From
this figure we see that for a given channel length the variance of
the mutual information varies monotonically with the SNR. This
increase is most significant for small values of $L$.

\begin{figure*}[!t]
\setcounter{mytempeqncnt}{\value{equation}}
\setcounter{equation}{31}
\begin{align} \label{eq:PhiFcn_defn_highSNR_SIMO}
\tilde{\varphi}(\rho_d) = \left\{
\begin{array}{ll}
 0 & \; \; \text{for} \; \; | \rho_d| = 0 \\
{\rm L_{i_2}}(1-|\rho_d|^2) - H^2(n-1) + 2 \sum_{b=1}^{n-1}
\frac{H(b-1)}{b} \\ + \sum_{b=1}^{n-1}
 \frac{1}{b} \biggl( \left( \frac{|\rho_d|^2-1}{|\rho_d|^2}\right)^b \ln (1 - |\rho_d|^2) - \sum_{t=0}^{b-2}
 \frac{ \left( \frac{|\rho_d|^2-1}{|\rho_d|^2}\right)^{t+1}}{b-t-1}  \biggr)   & \; \; \text{for} \; \;   0 < |\rho_d| < 1 \\
\psi'(n)
  & \; \; \text{for} \; \;  |\rho_d| = 1
\end{array}
\right.
\end{align}\setcounter{equation}{\value{mytempeqncnt}}
\hrulefill
\vspace*{4pt}
\end{figure*}

\subsection{Analysis at High SNR}

The following theorem presents a closed-form expression for the
variance of the mutual information of MIMO-OFDM in the high SNR
regime.  This result is simpler than the exact general variance
result given in Theorem \ref{th:exactVarAllSNR}, as it does not
involve any infinite series.

\begin{theorem} \label{th:VarHighSNR}
In the high SNR regime, the variance of the mutual information of
MIMO-OFDM systems is given by
\begin{align}
{\rm Var}^\infty(\mathcal{I}_{\rm ofdm}) &= (\log_2(e))^2 \Biggl(
\frac{2}{N^2}
 \sum_{d=1}^{N-1} (N - d) \tilde{\varphi}(\rho_d) \nonumber \\
 & \hspace*{-1cm} + \frac{1}{N} \sum_{t=0}^{m-1} \psi'(n-t) - \frac{N-1}{N} \biggl( \sum_{t=0}^{m-1} \psi (n -
t) \biggr)^2 \; \Biggr) \label{eq:VarHighSNRFinal}
\end{align}
where
\begin{align} \label{eq:PhiFcn_defn_highSNR}
\tilde{\varphi}(\rho_d) = \left\{
\begin{array}{ll}
\left( \sum_{t=0}^{m-1} \psi (n - t)
\right)^2 & \; \; \text{for} \; \; | \rho_d| = 0 \\
\frac{ \sum_{ r = 1 }^m \sum_{ s
= 1 }^m \det \bigl( \mathbf{\tilde{C}}_{r, s} (\rho_d) \bigr)}{\Gamma_m(n) \Gamma_m(m)}  & \; \; \text{for} \; \;   0 < |\rho_d| < 1 \\
\; \; \sum_{t=0}^{m-1} \psi'(n-t) \\ + \left( \sum_{t=0}^{m-1} \psi (n - t)
\right)^2
  & \; \; \text{for} \; \;  |\rho_d| = 1
\end{array}
\right.
\end{align}
where $\mathbf{\tilde{C}}_{r, s}(\rho_d)$ is an $m \times m$ matrix
with $(i,j)^{\rm th}$ element for the cases $i \neq r$ or $j \neq s$
given by
\begin{align} \label{eq:Ckell_defn_highSNR}
& \left(\mathbf{\tilde{C}}_{r, s}(\rho_d) \right)_{i,j}  \nonumber \\
& \hspace*{0.5cm} = \left\{
\begin{array}{ll}
\eta_{i,j} (1, \rho_d)  & \text{for} \; \; i \neq r, \; j \neq s \\
 \eta_{i,j} \left( \psi(z), \rho_d \right)  & \text{for} \; \;  i = r, \; j \neq s \\
|\rho_d|^{2(i-j)} \,   \eta_{j,i} \left( \psi(z), \rho_d \right) & \text{for} \; \;  i \neq r, \; j = s \\
\end{array}
\right.
\end{align}
and for the case $(i = r, \; j = s)$ by (\ref{eq:Ctilde_rs_Simple}) at the top of the page,
\addtocounter{equation}{1}
where $i' = \max(i,j)$ and $j' = \min(i,j)$.  Also, $\eta_{i,j}
(\cdot)$ is defined in (\ref{eq:EtaDefn}) in Theorem
\ref{th:exactVarAllSNR}, $\xi_{\cdot}(\cdot)$ is defined in
(\ref{eq:SijDefn}), $h(\cdot)$ is given by
\begin{align} \label{eq:suAsympt}
h(\rho_d) = \ln (1 - |\rho_d|^2) - \mathcal{K} \, ,
\end{align}
and $\mathcal{K} = 0.5772\ldots$ is the Euler-Mascheroni constant.
The function $H({\cdot})$ denotes the harmonic number
\begin{align} \label{eq:HarmDefn}
H(z) = \left\{
\begin{array}{ll}
\sum_{\ell = 1}^{z} \frac{1}{\ell} & \text{for} \; \; z > 0 \\
0  & \text{for} \; \; z = 0 \;
\end{array}
\right.
\end{align}
and $\psi (\cdot)$ is the  digamma function defined as \cite[eq.\
(6.3.2)]{abramowitz65}
\begin{align} \label{eq:digammaDefn}
\psi (n-t) = H (n-t-1) - \mathcal{K}
\end{align}
with first derivative $\psi'(\cdot)$ corresponding to the polygamma
function \cite[eq.\ (6.4.1)]{abramowitz65}.
\end{theorem}


\emph{Proof:} See Appendix \ref{app:ProofHighSNRVar}. \hfill
\interlinepenalty500 $\Box$

The following two corollaries present very simple high SNR variance
expressions for the special case of SIMO/MISO and SISO systems
respectively.


\begin{corollarybox} \label{corr:exactVarSIMOHighSNR}
The variance of the mutual information of SIMO/MISO-OFDM systems at
high SNR is given by
\begin{align} \label{eq:exactVarianceFinalMISOHighSNR}
& {\rm Var}^\infty(\mathcal{I}_{\rm ofdm}) \nonumber \\
 & \hspace*{0.5cm} =  (\log_2 (e))^2 \left(
\frac{2}{N^2} \sum_{d=1}^{N-1} (N-d) \, \tilde{\varphi} (\rho_d) +
\frac{\psi'(n)}{N} \right)
\end{align}
where $\tilde{\varphi}(\rho_d)$ is given by (\ref{eq:PhiFcn_defn_highSNR_SIMO}) at the top of the page,
with ${\rm L_{i_2}}(\cdot)$ denoting the \emph{dilogarithm} function
\cite[eq.\ (27.7.1)]{abramowitz65} .
\addtocounter{equation}{1}
\end{corollarybox}

\begin{corollarybox} \label{corr:exactVarSISOHighSNR}
The variance of the mutual information of SISO-OFDM systems at high
SNR is given by
\begin{align} \label{eq:exactVarianceFinalSISOHighSNR}
& {\rm Var}^\infty(\mathcal{I}_{\rm ofdm}) \nonumber \\
& \hspace*{0.5cm} =  (\log_2 (e))^2 \left(
\frac{2}{N^2} \sum_{d=1}^{N-1} (N-d) {\rm L_{i_2}}(1-|\rho_d|^2) +
\frac{\pi^2}{6 N}
 \right) .
\end{align}
\end{corollarybox}

\begin{figure} \centering
\includegraphics[width=\figwidth]{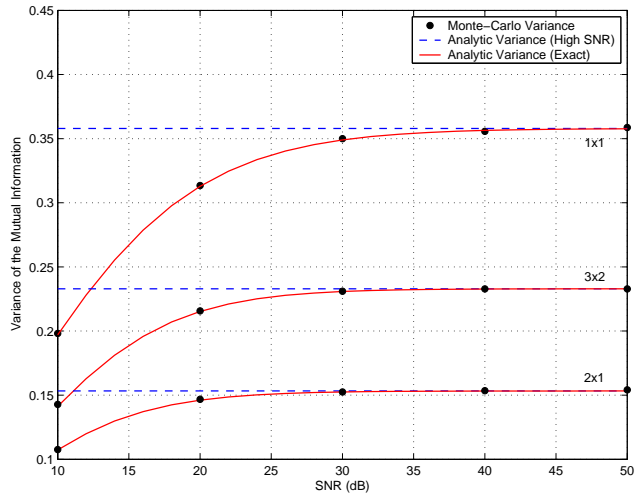}
\caption{Variance of the mutual information of MIMO-OFDM  for
different $N_t \times N_r$ antenna configurations, and for different
SNRs. The ``Analytic Variance (High SNR)'' lines are based on
(\ref{eq:VarHighSNRFinal}) for the $2 \times 3$ case,
(\ref{eq:exactVarianceFinalMISOHighSNR}) for the $1 \times 2$ case,
and (\ref{eq:exactVarianceFinalSISOHighSNR}) for the $1 \times 1$
case. The ``Analytic Variance (Exact)'' curves are based on
(\ref{eq:exactVarianceFinal}) for the $2 \times 3$ case,
(\ref{eq:exactVarianceFinalMISO}) for the $1 \times 2$ case, and
(\ref{eq:exactVarianceFinalSISO}) for the $1 \times 1$ case. $16$
subcarriers are considered, and the channel follows an $8$-path
uniform power delay profile.} \label{fig:var_Highsnr}
\end{figure}

It is important to note that the results in Theorem
\ref{th:VarHighSNR} and Corollaries \ref{corr:exactVarSIMOHighSNR}
and \ref{corr:exactVarSISOHighSNR} do not depend on the SNR. Therefore, a main insight which we can draw from these expressions is that the variance of the MIMO-OFDM mutual information converges to a deterministic limit as the SNR increases, which we have now quantified precisely.
This phenomenon is illustrated in Fig.\ \ref{fig:var_Highsnr}, where
we plot the variance of the MIMO-OFDM mutual information for
different $N_t \times N_r$ antenna configurations, and for different
SNRs. The ``Analytic Variance (High SNR)'' dashed lines are based on
(\ref{eq:VarHighSNRFinal}) for the $2 \times 3$ case,
(\ref{eq:exactVarianceFinalMISOHighSNR}) for the $1 \times 2$ case,
and (\ref{eq:exactVarianceFinalSISOHighSNR}) for the $1 \times 1$
case. The ``Analytic Variance (Exact)'' curves are based on
(\ref{eq:exactVarianceFinal}) for the $2 \times 3$ case,
(\ref{eq:exactVarianceFinalMISO}) for the $1 \times 2$ case, and
(\ref{eq:exactVarianceFinalSISO}) for the $1 \times 1$ case.
Monte-Carlo simulated variance curves are also presented for further
verification. We see that the results converge quickly in all cases.

\subsection{Analysis at Low SNR}

The following theorem presents a very simple closed-form expression
for the variance of the mutual information of MIMO-OFDM in the low
SNR regime.

\begin{theorem} \label{th:VarLowSNR}
In the low SNR regime, the variance of the mutual information of
MIMO-OFDM systems is given by
\begin{align}
{\rm Var}^0(\mathcal{I}_{\rm ofdm}) &= (\log_2(e))^2  \frac{\gamma^2
N_r}{N N_t } \left( 1 + 2 \sum_{d = 1}^{N-1} \frac{N-d}{N}
|\rho_d|^2 \right) \; . \label{eq:VarLowSNRFinal}
\end{align}
\end{theorem}

\emph{Proof:} See Appendix \ref{app:ProofLowSNRVar}. \hfill
\interlinepenalty500 $\Box$


The following corollary gives upper and lower bounds (as a function
of the frequency correlation coefficients) for the variance of the
MIMO-OFDM mutual information in the low SNR regime.

\begin{corollary}
In the low SNR regime, the variance of the mutual information of
MIMO-OFDM systems satisfies
\begin{align}\label{eq:corrLowSNRBounds}
\frac{1}{N} \hspace*{0.2cm}  \leq \hspace*{0.2cm} \frac{ {\rm Var}^0
(\mathcal{I}_{\rm ofdm})}{ {\rm Var}^0 (\mathcal{I}_{\rm flat})} =
\frac{1}{N} \left( 1 + 2 \sum_{d=1}^{N-1} \frac{N-d}{N} |\rho_d|^2
\right) \hspace*{0.2cm} \leq \hspace*{0.2cm} 1
\end{align}
where ${\rm Var}^0 (\mathcal{I}_{\rm flat})$ denotes the mutual
information variance for an i.i.d.\ flat-fading Rayleigh MIMO
channel.  The left-hand side is an equality for $|\rho_d| = 0$
(independent fading across all frequency subcarriers), and the
right-hand side is an equality for $|\rho_d| = 1$ (identical fading
across all subcarriers, i.e.\ flat-fading).
\end{corollary}


\emph{Proof:} The proof follows by using
\begin{align}
0 \; \leq \; | \rho_d| \; \leq \; 1
\end{align}
in (\ref{eq:VarLowSNRFinal}), and noting that
\begin{align}
{\rm Var}^0 (\mathcal{I}_{\rm flat}) = (\log_2(e))^2 \frac{ \gamma^2
N_r}{N_t} \; ,
\end{align}
which is found by directly setting $N = 1$ in
(\ref{eq:VarLowSNRFinal}).
 \hfill \interlinepenalty500 $\Box$

It is interesting to note from (\ref{eq:corrLowSNRBounds}) that in
the low SNR regime, the scaling of the MIMO-OFDM variance with
respect to the flat-fading variance depends only on the channel
delay profile, and is independent of the number of transmit and
receive antennas.


For the particular case of a uniform power delay profile (i.e. with
$\sigma_p^2 = 1/L$ for all $p = 0, \ldots, L-1$), we can obtain a
simple insightful expression for the variance ratio in
(\ref{eq:corrLowSNRBounds}), as given below.

\begin{corollary}
For a uniform power delay profile, (\ref{eq:corrLowSNRBounds})
becomes
\begin{align}\label{eq:corrLowSNRBounds_Uniform}
\frac{1}{N} \hspace*{0.2cm}  &\leq \hspace*{0.1cm} \frac{ {\rm Var}^0
(\mathcal{I}_{\rm ofdm})}{ {\rm Var}^0 (\mathcal{I}_{\rm flat})} \nonumber \\
& \hspace*{0.6cm} =
\frac{1}{N} \left( 1 + 2 \sum_{d=1}^{N-1} \frac{N-d}{N} \left(
\frac{ \sin \left( \frac{\pi d L}{N} \right)}{L \sin \left(
\frac{\pi d}{N} \right)} \right)^2 \right) \hspace*{0.1cm} \leq
\hspace*{0.1cm} 1 \;
\end{align}
where the left-hand side is an equality for $L = N$, and the
right-hand side is an equality for the case $L = 1$.
\end{corollary}


\emph{Proof:} The proof follows trivially from
(\ref{eq:corrLowSNRBounds}) after noting that the frequency
correlation-coefficients (\ref{eq:freqCorrCoef}) in this case can be
expressed as \cite{ko04}
\begin{align}
\rho_d = \frac{ \sin \left( \frac{\pi d L}{N} \right)}{L \sin \left(
\frac{\pi d}{N} \right)} e^{j \frac{\pi d}{ N}} \; .
\end{align}
 \hfill \interlinepenalty500 $\Box$

\begin{figure} \centering
\includegraphics[width=\figwidth]{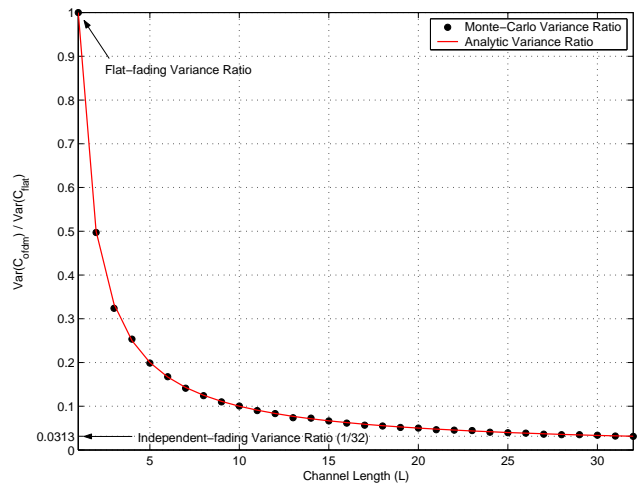}
\caption{Ratio of the MIMO-OFDM mutual information variance and the
flat-fading MIMO mutual information variance at low SNR, for
different channel lengths (uniform power delay profile). The
``Analytic Variance Ratio'' curve is based on
(\ref{eq:corrLowSNRBounds_Uniform}). A $2 \times 2$ system at $-25$
dB is considered with $32$ subcarriers.} \label{fig:var_LowSNRRatio}
\end{figure}

The summation in (\ref{eq:corrLowSNRBounds_Uniform}) is of a similar
type to that in \cite[eq.\ (60)]{moustakas06}, which gave an
asymptotic expression for the variance for large antenna numbers,
and involved the same squared-ratio terms. As mentioned in
\cite{moustakas06}, as $L$ increases, the ratio becomes more peaked
as a function of $d$, thereby decreasing the overall sum. Thus, from
(\ref{eq:corrLowSNRBounds_Uniform}) we see that the variance of the
mutual information varies inversely with the channel delay spread in
the low SNR regime. This agrees with previous observations seen via
simulation studies in \cite{bolcskei02}, and for the regime of large
antenna numbers in \cite{moustakas06}.  These results are further
corroborated in Fig.\ \ref{fig:var_LowSNRRatio}, where
(\ref{eq:corrLowSNRBounds_Uniform}) is plotted as a function of the
channel length $L$.

\section{Outage Approximation of MIMO-OFDM Based
Spatial-Multiplexing} \label{sec:outageApprox}

We now use the analytic expressions from the previous section to
present and investigate approximations for the distribution of
mutual information.  We then use the approximations to estimate
outage capacity.

Unless otherwise stated, for all results in this section we model
the channel according to the exponential power delay profile
\cite{muller-weinfurtner02}
\begin{align}
\sigma^2_p = \left\{
\begin{array}{ll}
\frac{1 - e^{-1/K_{\rm exp}}}{1 - e^{-L/K_{\rm exp}}} e^{-p/K_{\rm exp}} & \; \; \text{for} \; \;  0 \leq p < L \\
0  & \; \; \text{otherwise}  \\
\end{array}
\right.
\end{align}
where $K_{\rm exp}$ is a parameter which characterizes the rate of
decay of the power delay profile as a function of $p$, and is
loosely related to the rms delay spread \cite{muller-weinfurtner02}.

\begin{figure} \centering
\includegraphics[width=\figwidth]{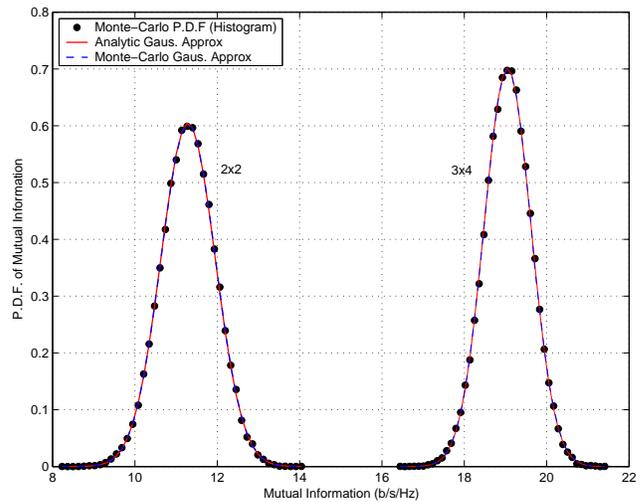}
\caption{P.d.f. of MIMO-OFDM mutual information for different $N_t
\times N_r$ antenna configurations. The ``Analytic Gaus. Approx.''
curves are based on the exact mean formula (\ref{eq:MeanFlatFading})
and exact variance formula (\ref{eq:exactVarianceFinal}). $64$
subcarriers is considered, with SNR of $20$ dB. The channel follows
an $8$-path exponential power delay profile with $K_{\rm exp} = 4$.}
\label{fig:pdf_diffAnt}
\end{figure}

\subsection{Gaussian and Gamma Approximations}

We first investigate the accuracy of a Gaussian approximation for
various system configurations and channel scenarios.

Fig.~\ref{fig:pdf_diffAnt} presents the analytical Gaussian
approximation for the MIMO-OFDM mutual information p.d.f. based on
the exact mean and variance expressions in (\ref{eq:MeanFlatFading})
and (\ref{th:exactVarAllSNR}) respectively, as well as empirically
generated p.d.f.s (Monte-Carlo histogram), for different antenna
configurations. A $64$-subcarrier system is considered with SNR of
$20$ dB. We see that the analytic curves match the true distribution
almost perfectly for both antenna configurations.  We also present
curves for a simulation based Gaussian approximation (based on the
mean and variance of the Monte-Carlo generated histograms) for
further verification. Note that these curves are indistinguishable
from our new analytical Gaussian approximation curves.

\begin{figure} \centering
\includegraphics[width=\figwidth]{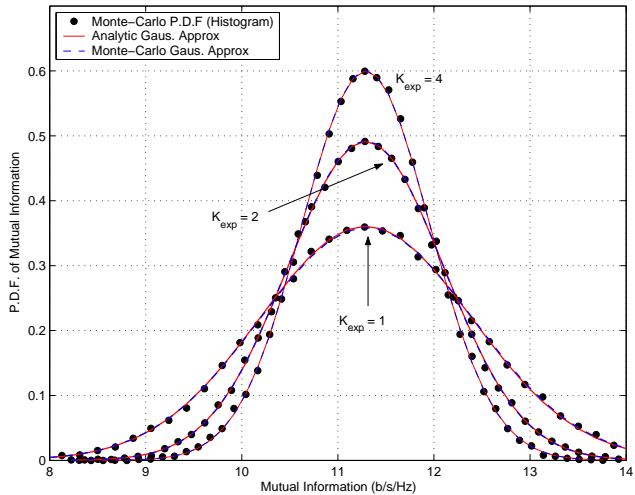}
\caption{P.d.f. of MIMO-OFDM mutual information for different rms
delay spreads (channels follow an $8$-path exponential power delay
profile, with different $K_{\rm exp}$). The ``Analytic Gaus.
Approx.'' curves are based on the exact mean formula
(\ref{eq:MeanFlatFading}) and exact variance formula
(\ref{eq:exactVarianceFinal}). $16$ subcarriers is considered, with
SNR of $20$ dB.} \label{fig:pdf_diffDelaySp}
\end{figure}

Fig.~\ref{fig:pdf_diffDelaySp} compares the analytical Gaussian
approximation with empirically-generated p.d.f.\ curves, for
different channel rms delay spreads.  Again we see that the analytic
Gaussian approximation is accurate in all cases.  Moreover, we see a
significant reduction in the variance of the mutual information as
the rms delay spread increases (i.e.\ as $K_{\rm exp}$ increases).
Again note that the Monte-Carlo Gaussian approximation is
indistinguishable from our new analytical Gaussian approximation
curves.

\begin{figure} \centering
\includegraphics[width=\figwidth]{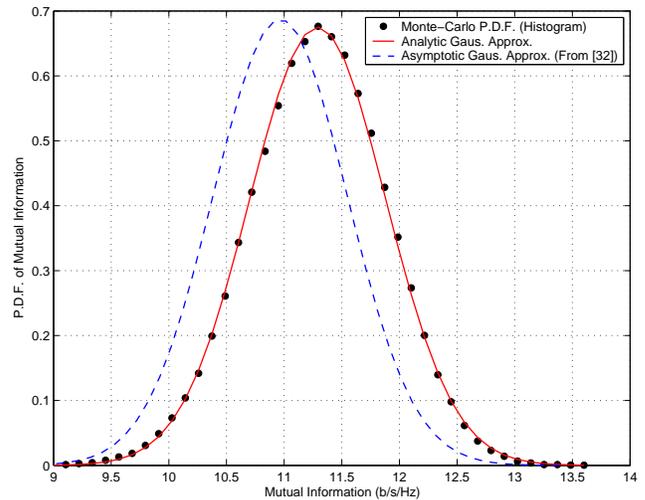}
\caption{P.d.f. of MIMO-OFDM mutual information. The ``Analytic
Gaus. Approx.'' curve is based on the exact mean formula
(\ref{eq:MeanFlatFading}) and exact variance formula
(\ref{eq:exactVarianceFinal}). The ``Asymptotic Gaus. Approx. (From
\cite{moustakas06})'' curve is based on \cite[eqs.\ (59) and
(60)]{moustakas06}. $2 \times 2$ antennas and $32$ subcarriers are
considered, with $20$ dB SNR.  The channel follows an $8$-path
uniform power delay profile.} \label{fig:Compare_Moustakas_Diffsnr}
\end{figure}

\begin{figure} \centering
\includegraphics[width=\figwidth]{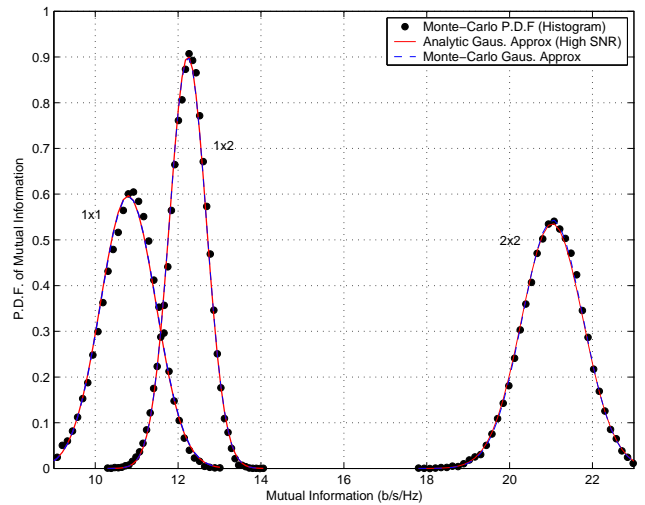}
\caption{P.d.f. of MIMO-OFDM mutual information at high SNR.  The
``Analytic Gaus. Approx (High SNR)'' curves are based on the high
SNR mean formula \cite[Theorem 2]{grant02}, and the variance formula
(\ref{eq:VarHighSNRFinal}) for the $2 \times 2$ case,
(\ref{eq:exactVarianceFinalMISOHighSNR}) for the $1 \times 2$ case,
and (\ref{eq:exactVarianceFinalSISOHighSNR}) for the $1 \times 1$
case. $16$ subcarriers is considered, with SNR of $35$ dB. The
channel follows an $8$-path exponential power delay profile with
$K_{\rm exp} = 4$. } \label{fig:pdf_diffAnt_HighSNR}
\end{figure}

Fig.~\ref{fig:Compare_Moustakas_Diffsnr} compares our new analytic
Gaussian approximation with the asymptotic Gaussian approximation
previously derived in \cite{moustakas06}; formally derived under the
assumption of asymptotically large antenna numbers.  To our
knowledge, this is the only other comparable analytical result in
the literature which applies for arbitrary-length
frequency-selective MIMO channels. In the figure, we consider a $2
\times 2$ system at $20$ dB SNR. The channel has a uniform
power-delay profile; for which simple approximations to the mean and
variance of the mutual information were explicitly presented in
\cite[eqs.\ (59) and (60)]{moustakas06}. Clearly, although the
approximation in \cite{moustakas06} was shown to be quite accurate
for some practical scenarios,
Fig.~\ref{fig:Compare_Moustakas_Diffsnr} shows that our analytic
Gaussian approximation is more accurate (although it is only shown
for $32$ subcarriers, the same observation has been made for all
systems investigated).

Fig.~\ref{fig:pdf_diffAnt_HighSNR} presents the distribution of the
mutual information at high SNRs, comparing MIMO, SIMO, and SISO
systems. The analytic Gaussian approximation curves are based on a
high SNR mean formula from \cite[Theorem 2]{grant02}, and the high
SNR variance formula (\ref{eq:VarHighSNRFinal}) for the MIMO case,
(\ref{eq:exactVarianceFinalMISOHighSNR}) for the SIMO case, and
(\ref{eq:exactVarianceFinalSISOHighSNR}) for the SISO case.  We see
that the analytic Gaussian approximation is accurate in all cases.
Again note that the Monte-Carlo Gaussian approximation is
indistinguishable from our new analytical Gaussian approximation
curves.

\begin{figure} \centering
\includegraphics[width=\figwidth]{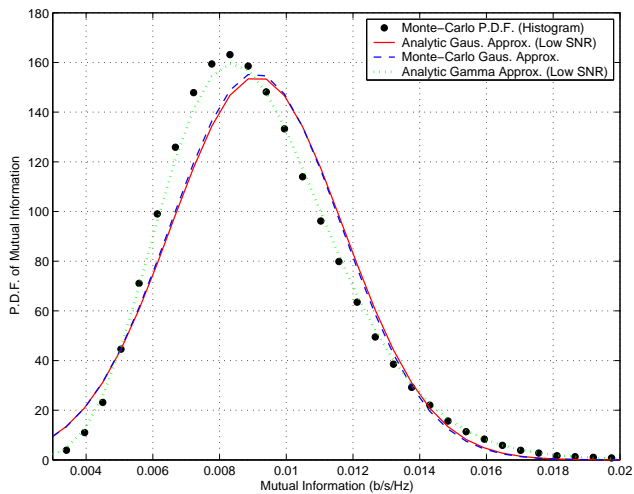}
\caption{P.d.f. of MIMO-OFDM mutual information at low SNR.  The
``Analytic Gaus. Approx (Low SNR)'' and ``Analytic Gamma Approx (Low
SNR)'' curves are based on the low SNR mean formula obtained by
combining (\ref{eq:TaylorFirstOrder}) and (\ref{eq:traceIdent}), and
the low SNR variance formula (\ref{eq:VarLowSNRFinal}).  A $2 \times
2$ system with $32$ subcarriers is considered, with SNR of $-25$ dB.
The channel follows a $4$-path exponential power delay profile with
$K_{\rm exp} = 2$. } \label{fig:pdf_LowSNR_Gamma}
\end{figure}

Fig.~\ref{fig:pdf_LowSNR_Gamma} presents the distribution of the
mutual information at low SNRs.  The analytic Gaussian approximation
curve is generated based on the low SNR mean formula obtained by
combining (\ref{eq:TaylorFirstOrder}) and (\ref{eq:traceIdent}), and
the low SNR variance formula (\ref{eq:VarLowSNRFinal}). In this case
we see that a Gaussian distribution no longer accurately predicts
the mutual information p.d.f. This can be explained by examining
(\ref{eq:TaylorFirstOrder}), where we see that at low SNRs the
mutual information for each subcarrier is a function of ${\rm
tr}\left( \mathbf{H}_k \mathbf{H}_k^\dagger \right)$, which for
i.i.d.\ Rayleigh fading is $\sim \chi_{2 N_r N_t}^2$.  Hence, the
overall mutual information (\ref{eq:meanMI}) is distributed as the
sum of $N$ correlated $\chi_{2 N_r N_t}^2$ random variables which
(for small $N$), is clearly quite different to Gaussian.

Motivated by this observation, we propose to approximate the mutual
information p.d.f.\ at low SNR with a Gamma distribution. Note that
a Gamma approximation was previously considered in the context of
flat-fading channels in \cite{kang03_Gamma}. The Gamma p.d.f.\ is
given by
\begin{align}
f(x) = \frac{ \theta^r x^{r-1} e^{-\theta x}}{\Gamma(r)} \; ,
\hspace*{1cm} x \geq 0
\end{align}
where $r$ is the shape parameters and $\theta$ is the scale
parameter. By matching the first two moments, a Gamma approximation
for the mutual information p.d.f.\ of MIMO-OFDM is obtained by
evaluating
\begin{align}
r = \frac{ E \left[ \mathcal{I}_{\rm ofdm} \right]}{ {\rm Var}\left(
\mathcal{I}_{\rm ofdm} \right)}
\end{align}
and
\begin{align} \theta = \frac{ E^2
\left[ \mathcal{I}_{\rm ofdm} \right]}{ {\rm Var}\left(
\mathcal{I}_{\rm ofdm} \right)} \; .
\end{align}
This analytic Gamma approximation is plotted in
Fig.~\ref{fig:pdf_LowSNR_Gamma}, based on the same low SNR analytic
mean and variance formulas as used for the low SNR Gaussian
approximation above. We clearly see that the Gamma approximation is
much more accurate than the Gaussian approximation in this low SNR
regime, and follows the simulated p.d.f.\ very closely.

\begin{figure} \centering
\includegraphics[width=\figwidth]{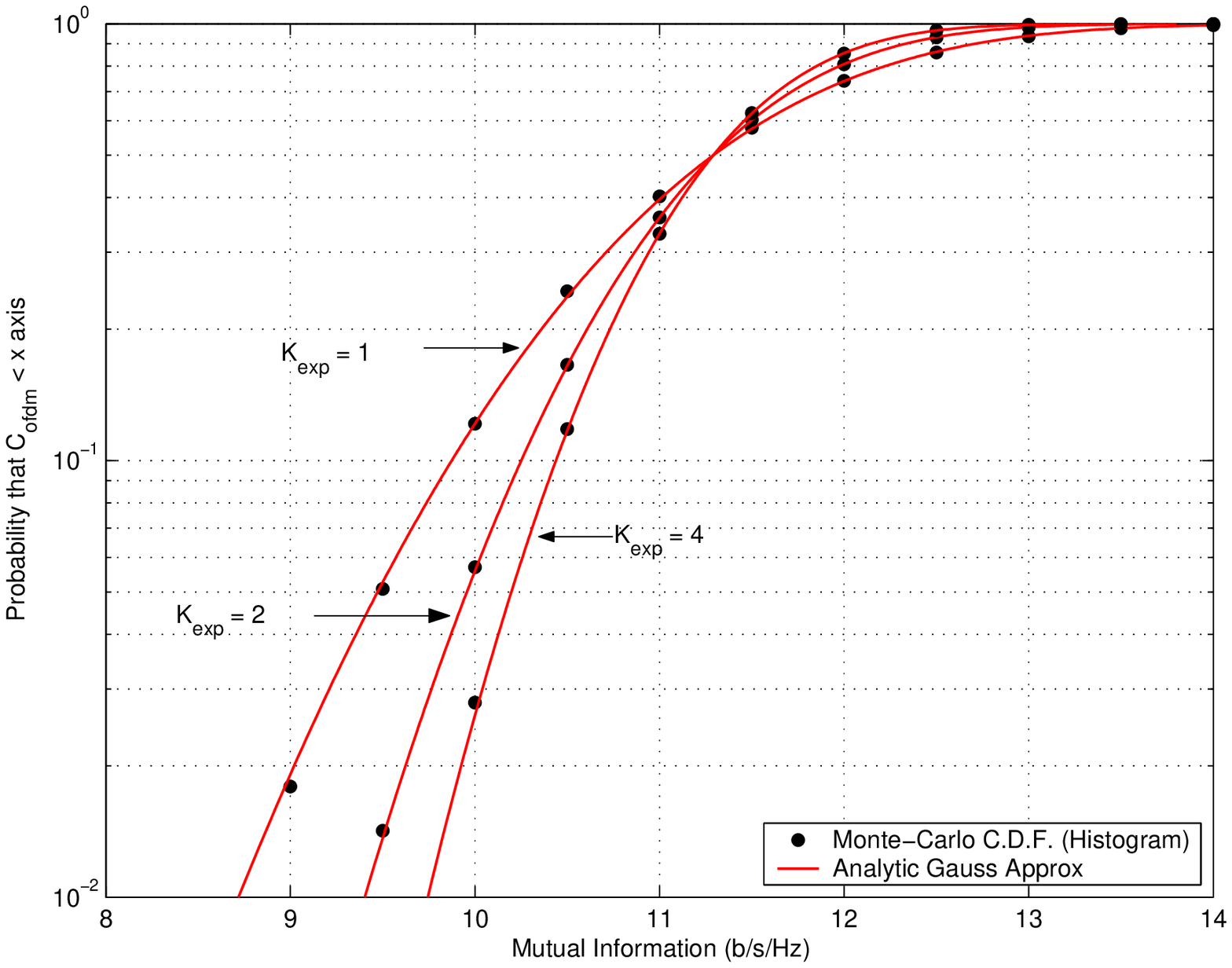}
\caption{C.d.f. of MIMO-OFDM mutual information for different rms
delay spreads (channels follow an $8$-path exponential power delay
profile, with different $K_{\rm exp}$).  The ``Analytic Gaus.
Approx'' curves are based on the exact mean formula
(\ref{eq:MeanFlatFading}) and exact variance formula
(\ref{eq:exactVarianceFinal}). A $2 \times 2$ system with $16$
subcarriers is considered, with SNR of $20$ dB. } \label{fig:outage}
\end{figure}

\subsection{Outage Capacity}

The outage capacity $\mathcal{I}_{out,q}$ is defined as the maximum
information rate guaranteed to be supported for $100(1 - q)\%$ of
the channel realizations\footnote{Strictly speaking, computing the outage capacity would require performing a numerical optimization over all possible input distributions, as discussed in \cite{teletar99}.  Here however, we adopt a common slight abuse of terminology, and use the term outage capacity to denote the outage rate for the case of OFDM-based spatial multiplexing systems with equal power Gaussian inputs.}, ie.\
\begin{align} \label{eq:OutageCDF}
P (\mathcal{I}_{\rm ofdm} \leq \mathcal{I}_{{\rm out}, q}) = q
\end{align}
where $q$ denotes the outage probability, and is thus directly
obtained by inverting the c.d.f.\ of $\mathcal{I}_{\rm ofdm}$.
If the distribution of the mutual information is Gaussian, then the
outage capacity can be computed from the derived mean and variance
as \cite[eq.\ (26)]{barriac04}
\begin{equation}
\mathcal{I}_{{\rm out},q} =  E \left[ \mathcal{I}_{\rm ofdm} \right]
- \sqrt{{\rm Var}(\mathcal{I}_{\rm ofdm})} Q^{-1}(q)
\end{equation}
where $Q(\cdot)$ is the Gaussian $Q$-function.

Fig.\ \ref{fig:outage} plots the outage probability for channels
with different rms delay spreads. The ``Analytic Gaus Approx''
curves are generated by approximating the c.d.f.\ in
(\ref{eq:OutageCDF}) as a Gaussian distribution, and using the exact
mean and variance formulas in (\ref{eq:MeanFlatFading}) and
(\ref{th:exactVarAllSNR}) respectively. Clearly this analytic
Gaussian approximation matches closely with the empirically
generated c.d.f.\ (Monte-Carlo histogram) in all cases. Moreover, we
see that for outage probabilities of practical interest (e.g.\ $q =
1\%$), increasing the rms delay spread can yield a significant
improvement in outage capacity.

\section{Conclusions}

This paper has considered the mutual information distribution of
frequency-selective MIMO channels, in the context of OFDM-based
spatial multiplexing systems.  Exact closed-form expressions were
presented for the mutual information variance, applying for arbitrary finite
system and channel parameters. These results were used to provide
accurate analytical approximations for the distribution of mutual
information, and the outage capacity. We observed that for most
scenarios a Gaussian approximation is accurate, while also noting
that for low SNR a Gamma approximation yielded even higher accuracy.

\begin{appendices}

\section{Proof of Theorem \ref{th:exactVarAllSNR}}
\label{app:ProofExactVar}

\emph{Proof:}
By definition, the variance of the mutual information is given by
\begin{align}
{\rm Var}(\mathcal{I}_{\rm ofdm}) = E \left[ \mathcal{I}^2_{\rm
ofdm} \right] - E^2 \left[ \mathcal{I}_{\rm ofdm} \right] \; .
\end{align}

Noting that $E \left[ \mathcal{I}_{\rm ofdm} \right] = E \left[
\mathcal{I}_{\rm flat} \right]$,
and using (\ref{eq:OFDMMi}), we have
\begin{align}
{\rm Var}(\mathcal{I}_{\rm ofdm}) &= E \left[ \frac{1}{N^2}
\sum_{k=0}^{N-1} \sum_{\ell = 0}^{N-1} \mathcal{I}_k
\mathcal{I}_\ell \right] - E^2 \left[ \mathcal{I}_{\rm flat}
\right] \nonumber \\
& \hspace*{-1.5cm} = \frac{1}{N^2} \left( \sum_{k=0}^{N-1} \sum_{\ell = 0, \ell \neq
k}^{N-1} E \left[ \mathcal{I}_k \mathcal{I}_\ell \right] + \sum_{k =
0}^{N-1} E \left[ \mathcal{I}_k^2 \right] \right) - E^2 \left[
\mathcal{I}_{\rm flat}
\right] \nonumber \\
& \hspace*{-1.5cm} = \frac{1}{N^2} \left( \sum_{k=0}^{N-1} \sum_{\ell = 0, \ell \neq
k}^{N-1} E \left[ \mathcal{I}_k \mathcal{I}_\ell \right] \right) +
\frac{1}{N} E \left[ \mathcal{I}_{\rm flat}^2 \right] - E^2 \left[
\mathcal{I}_{\rm flat} \right] \label{eq:VarEqn}
\end{align}
where $\mathcal{I}_{\rm flat}$ denotes the mutual information of a
flat-fading channel. Note that the last line followed by noting
that, under the assumptions in Section \ref{sec:ChanModel}, the
channel statistics for each subcarrier (and therefore, the mutual
information statistics) are identical \cite{bolcskei02}, and
moreover, these statistics are equal to that of a flat-fading
i.i.d.\ Rayleigh channel. The first and second moments of the mutual
information for flat-fading channels has been previously derived in
terms of incomplete gamma functions in \cite[eqs. (29) and
(31)]{kang04}. Using \cite[eq.\ (6.5.9)]{abramowitz65}, we perform
some basic manipulations to express these results in alternative
simplified forms as follows
\begin{align} \label{eq:MeanFlatFading}
& E \left[ \mathcal{I}_{\rm flat} \right] = \frac{
\log_2(e)}{\Gamma_m(n)
\Gamma_m(m)} \sum_{r=1}^m \det \left( \mathbf{A}_r \right) \\
 \label{eq:SecMomFlatFading}
& E \left[ \mathcal{I}_{\rm flat}^2 \right] = \frac{ (\log_2 (e))^2
}{\Gamma_m(n) \Gamma_m(m)} \sum_{r=1}^m \sum_{s = 1}^m \det \left(
\mathbf{B}_{r,s} \right)
\end{align}
where $\mathbf{A}_r$ and $\mathbf{B}_{r,s}$ are defined in
(\ref{eq:Ar_defn}) and (\ref{eq:Brs_defn}) respectively.

The challenge is to evaluate the cross-correlation of the mutual
information across frequency subcarriers $E \left[ \mathcal{I}_k
\mathcal{I}_\ell \right]$ which, using (\ref{eq:logDet}), is given
by
\begin{align} \label{eq:capacityCrossCorr}
E \left[ \mathcal{I}_k \mathcal{I}_\ell \right] &= E \Biggl[ \log_2
\det \left( \mathbf{I}_{N_r} + \frac{\gamma}{N_t} \mathbf{H}_k
\mathbf{H}_k^\dagger \right) \nonumber \\
& \hspace*{1cm} \times \log_2 \det \left( \mathbf{I}_{N_r} +
\frac{\gamma}{N_t} \mathbf{H}_\ell \mathbf{H}_\ell^\dagger \right)
\Biggl] \nonumber \\
&= E \left[ \sum_{i=1}^m \log_2 \left( 1
+ \frac{\gamma}{N_t} \lambda_i \right) \sum_{j=1}^m \log_2 \left( 1
+ \frac{\gamma}{N_t} \omega_j \right) \right]
\end{align}
where $\mathbf{\underline{\lambda}} = \left\{ \lambda_i
\right\}_{i=1}^m$ and $\mathbf{\underline{\omega}} = \left\{
\omega_i \right\}_{i=1}^m$ are the non-zero eigenvalues of
$\mathbf{H}_k \mathbf{H}_k^\dagger$ and $\mathbf{H}_\ell
\mathbf{H}_\ell^\dagger$ respectively. Defining
\begin{align}
\alpha (x) = \log_2 \left( 1 + \frac{ \gamma }{N_t} \, x \right)
\end{align}
we have
\begin{align} \label{eq:IkIellSum}
E \left[ \mathcal{I}_k \mathcal{I}_\ell \right] &=E \left[
\sum_{i=1}^m \sum_{j=1}^m \alpha(\lambda_i) \alpha (\omega_j )
\right] \nonumber \\
&=  \, \sum_{i=1}^m \sum_{j=1}^m E \left[ \alpha(\lambda_i) \alpha (\omega_j )
\right] .
\end{align}
Now, to evaluate the expectations in (\ref{eq:IkIellSum}), we first simplify the problem by exploiting the symmetry with respect to the $\lambda_i$s and $\omega_j$s.  To this end, let $\lambda$ and $\omega$ be randomly (uniformly) chosen
eigenvalues from $\mathbf{\underline{\lambda}}$ and
$\mathbf{\underline{\omega}}$ respectively.  Then clearly
\begin{align} \label{eq:Prtmp}
{\rm Pr} \left( \lambda = \lambda_i, \omega =
\omega_j \right) = \frac{1}{m^2},
\end{align}
for any given $i \in \{1,\ldots,m\}, j \in \{1,\ldots,m\}$.  Hence, we can also write
\begin{align}
E \left[ \alpha(\lambda) \alpha(\omega) \right] &= \sum_{i=1}^m \sum_{j=1}^m {\rm Pr}(\lambda = \lambda_i, \omega = \omega_j) \nonumber \\
  & \hspace*{1cm} \times E \left[ \alpha (\lambda) \alpha (\omega) | \lambda = \lambda_i, \omega = \omega_j  \right] \; \nonumber \\
&= \frac{1}{m^2}
\sum_{i=1}^m \sum_{j=1}^m E \left[ \alpha (\lambda_i) \alpha
(\omega_j) \right] \; .  \label{eq:Elamom}
\end{align}
where the second line follows from (\ref{eq:Prtmp}).
Therefore by directly comparing (\ref{eq:Elamom}) with (\ref{eq:IkIellSum}) it follows that
\begin{align} \label{eq:CrossCorrExp}
E \left[ \mathcal{I}_k \mathcal{I}_\ell \right] = m^2 E \left[
\alpha(\lambda) \alpha(\omega) \right] \; .
\end{align}
We point out that the simplification from (\ref{eq:IkIellSum}) to (\ref{eq:CrossCorrExp}) is particularly important, since in order to evaluate the expectation in (\ref{eq:CrossCorrExp}), clearly we only require the distribution of a pair of arbitrarily-selected eigenvalues, $\lambda$ and $\omega$. This turns out to be much more convenient than dealing with the distributions of the individual pairs of ordered eigenvalues, ie.\ $\lambda_i$ and $\omega_j$, required to directly evaluate (\ref{eq:IkIellSum}).

\begin{figure*}[!t]
\setcounter{mytempeqncnt}{\value{equation}}
\setcounter{equation}{59}
\begin{align} \label{eq:Dunline}
\left(\mathbf{\underline{D}}_{r, s} \right)_{i,j} = \left\{
\begin{array}{ll}
a(i,j) \defeq \int_0^\infty \int_0^\infty
\lambda^{\frac{\tau}{2}+i-1} \omega^{\frac{\tau}{2}+j-1} e^{-\frac{
\lambda}{1-|\rho_d|^2} } e^{ -\frac{ \omega}{1-|\rho_d|^2} } \\
\hspace*{4cm} \times I_{\tau} \left( \frac{ 2 |\rho_d|
}{1-|\rho_d|^2} \sqrt{ \lambda
\omega } \right) \alpha(\lambda) \alpha(\omega) {\rm d}\lambda {\rm d}\omega & \text{for} \; \; i = r, \; j = s \\
b(i,j) \defeq \int_0^\infty \frac{ \Gamma(\tau+j) \, |\rho_d|^\tau
}{\left(1 - |\rho_d|^2\right)^{-j}} e^{-\lambda} \lambda^{\tau + i -
1} \sum_{t=0}^{j-1} \binom{j-1}{t} \left( \frac{ |\rho_d|^2 \lambda
}{1 -
|\rho_d|^2} \right)^t \frac{ \alpha(\lambda) }{ (\tau + t)! }  {\rm d}\lambda & \text{for} \; \;  i = r, \; j \neq s \\
c(i,j) \defeq \int_0^\infty \frac{\Gamma(\tau+i)
|\rho_d|^\tau}{\left(1 - |\rho_d|^2\right)^{-i}} e^{-\omega}
\omega^{\tau + j - 1} \sum_{t=0}^{i-1} \binom{i-1}{t} \left( \frac{
|\rho_d|^2 \omega }{1 - |\rho_d|^2} \right)^t \frac{ \alpha(\omega) }{ (\tau + t)! }   {\rm d}\omega & \text{for} \; \;  i \neq r, \; j = s \\
\frac{\Gamma(\tau+j) \, |\rho_d|^\tau}{\left(1 -
|\rho_d|^2\right)^{-j}} \sum_{t=0}^{j-1} \binom{j-1}{t} \left(
\frac{ |\rho_d|^2 }{1 - |\rho_d|^2} \right)^t \frac{ (\tau + i + t -
1)! }{ (\tau + t)! } & \text{for} \; \; i \neq r, \; j \neq s \\
\end{array}
\right.
\end{align}
\setcounter{equation}{\value{mytempeqncnt}}
\hrulefill
\vspace*{4pt}
\end{figure*}

\begin{figure*}[!t]
\setcounter{mytempeqncnt}{\value{equation}}
\setcounter{equation}{67}
\begin{align}
\left(\mathbf{D}_{r,s} \left( \lambda, \omega \right) \right)_{i,j}
= \left\{
\begin{array}{ll}
\lambda^{\frac{\tau}{2}+i-1} \omega^{\frac{\tau}{2}+j-1} e^{-\frac{
\lambda}{1-|\rho_{d}|^2} } e^{ -\frac{ \omega}{1-|\rho_{d}|^2} }
I_{\tau} \left( \frac{ 2 |\rho_{d}| }{1-|\rho_{d}|^2} \sqrt{ \lambda
\omega } \right) & \text{for} \; \; i = r, \; j = s \\
\frac{ \Gamma(\tau+j) \, |\rho_{d}|^\tau }{\left(1 -
|\rho_{d}|^2\right)^{-j}} e^{-\lambda} \lambda^{\tau + i - 1}
\sum_{t=0}^{j-1} \binom{j-1}{t} \left( \frac{ |\rho_{d}|^2 \lambda
}{1 -
|\rho_{d}|^2} \right)^t \frac{ 1 }{ (\tau + t)! }  & \text{for} \; \;  i = r, \; j \neq s  \\
\frac{\Gamma(\tau+i) |\rho_{d}|^\tau}{\left(1 -
|\rho_{d}|^2\right)^{-i}} e^{-\omega} \omega^{\tau + j - 1}
\sum_{t=0}^{i-1} \binom{i-1}{t} \left( \frac{ |\rho_{d}|^2 \omega
}{1 -
|\rho_{d}|^2} \right)^t \frac{ 1 }{ (\tau + t)! } & \text{for} \; \;  i \neq r, \; j = s \\
\frac{\Gamma(\tau+j) \, |\rho_{d}|^\tau}{\left(1 -
|\rho_{d}|^2\right)^{-j}} \sum_{t=0}^{j-1} \binom{j-1}{t} \left(
\frac{ |\rho_{d}|^2 }{1 - |\rho_{d}|^2} \right)^t \frac{ (\tau + i +
t - 1)! }{ (\tau + t)! } & \text{for} \; \;  i \neq r, \; j \neq s
\end{array}
\right.
\label{eq:DUnlineCF}
\end{align}
\setcounter{equation}{\value{mytempeqncnt}}
\hrulefill
\vspace*{4pt}
\end{figure*}

The joint p.d.f.\ of $\lambda$ and $\omega$ is presented in Lemma \ref{lem:ArbEV}
in Appendix \ref{sec:App1_ArbEv}.  From this lemma we see that
$f(\lambda,w)$, and correspondingly $E \left[ \mathcal{I}_k
\mathcal{I}_\ell \right]$ in (\ref{eq:CrossCorrExp}), only depends
on $k$ and $\ell$ through their absolute difference, i.e.\ since
$f(\lambda,w)$ only depends on $k$ and $\ell$ via $| \rho_{k-\ell}
|$, and from (\ref{eq:freqCorrCoef})
\begin{align}
| \rho_{k-\ell} | = | \rho^*_{\ell-k} | = | \rho_{\ell-k} | \; .
\end{align}
Therefore the left-hand summation in (\ref{eq:VarEqn}) can be
written as
\begin{align} \label{eq:ACFToneDiff}
\sum_{k=0}^{N-1} \sum_{\ell = 0, \ell \neq k}^{N-1} E \left[
\mathcal{I}_k \mathcal{I}_\ell \right] = 2 \sum_{d=1}^{N-1} (N - d)
E \left[ \mathcal{I}_0 \mathcal{I}_{d} \right] \; .
\end{align}
Note that for subcarrier spacings $d$ for which the frequency
matrices are independent (i.e.\ $\rho_d = 0$) or completely
correlated (i.e.\ $\rho_d = 1$), the expectations in
(\ref{eq:ACFToneDiff}) are evaluated trivially as
\begin{align} \label{eq:Cd_Extreme}
&E \left[ \mathcal{I}_0 \mathcal{I}_{d} \right] = E^2 \left[
\mathcal{I}_{\rm flat} \right] \; ,
\hspace*{1cm} \; \;  \rho_{d} = 0 \nonumber \\
& E \left[ \mathcal{I}_0 \mathcal{I}_{d} \right] = E \left[
\mathcal{I}^2_{\rm flat} \right] , \; \hspace*{1cm} \; \; \rho_{d} =
1 \; .
\end{align}
For the case $0 < |\rho_d| < 1$ such a direct evaluation is not
possible, and we use (\ref{eq:CrossCorrExp}) in Lemma
\ref{lem:ArbEV} and (\ref{eq:arbUnorderedPdf}) to evaluate the
expectations in (\ref{eq:ACFToneDiff}) as follows
\begin{align} \label{eq:ACF_capacity_start}
E \left[ \mathcal{I}_0 \mathcal{I}_{d} \right] &= m^2 \int_0^\infty
\int_0^\infty \frac{ \alpha(\lambda)\alpha(\omega)  |\rho_d|^{-m
(n-1)}}{\Gamma_m(n) \Gamma_m(m) m^2 \left(1 - |\rho_d|^2 \right)^m } \nonumber \\
& \times \sum_{ r = 1 }^m \sum_{ s = 1 }^m \det \left( \mathbf{D}_{r, s}
\left( \lambda, \omega
\right) \right) {\rm d}\lambda {\rm d}\omega \nonumber \\
&= \frac{  |\rho_d|^{-m (n-1)}}{\Gamma_m(n) \Gamma_m(m) \left(1 -
|\rho_d|^2 \right)^m } \sum_{ r = 1 }^m \sum_{ s = 1 }^m \det \left(
\mathbf{\underline{D}}_{r, s} \right)
\end{align}
for $0 < |\rho_{d}| < 1$,
where $\mathbf{\underline{D}}_{r, s}$ is an $m \times m$ matrix with
$(i,j)^{\rm th}$ element defined in (\ref{eq:Dunline}) at the top of the page.
\addtocounter{equation}{1}

\begin{figure*}[!t]
\setcounter{mytempeqncnt}{\value{equation}}
\setcounter{equation}{75}
\begin{align}
\left(\tilde{\mathbf{D}}_{\alpha,\beta} \left( \lambda, \omega
\right) \right)_{i,j} = \left\{
\begin{array}{ll}
 a(\lambda, \omega, \alpha_i, \beta_j) & \text{for} \; \; i = 1, \; j = 1 \\
 b(\lambda, \alpha_i, \beta_j) \defeq \int_0^\infty a(\lambda, \omega_j, \alpha_i, \beta_j) {\rm d}\omega_j  & \text{for} \; \;  i = 1, \; j \neq 1 \\
 c(\omega, \alpha_i, \beta_j) \defeq \int_0^\infty a(\lambda_i, \omega, \alpha_i, \beta_j) {\rm d}\lambda_i & \text{for} \; \;  i \neq 1, \; j = 1 \\
 d(\alpha_i, \beta_j) \defeq \int_0^\infty \int_0^\infty a(\lambda_i, \omega_j, \alpha_i, \beta_j) {\rm d}\lambda_i {\rm d}\omega_j & \text{for} \; \; i \neq 1, \; j \neq 1 \\
\end{array}
\right.
\label{eq:DtildeCF}
\end{align}
\setcounter{equation}{\value{mytempeqncnt}}
\hrulefill
\vspace*{4pt}
\end{figure*}

Using the identity \cite{alfano04_2}
\begin{align} \label{eq:intIdentExpInt}
\int_0^\infty \ln \left( 1 + \alpha \lambda \right) \lambda^{q-1}
e^{-b\lambda} {\rm d}\lambda = \Gamma(q) e^{b/\alpha} b^{-q}
\sum_{h=1}^q {\rm E}_h \left( \frac{b}{\alpha} \right)
\end{align}
we can evaluate $b(i,j)$ and $c(i,j)$ in closed-form as
\begin{align} \label{eq:bijFinal}
b(i,j) &= \frac{ \log_2(e) e^{N_t/\gamma}  \Gamma(\tau + j) \,
|\rho_d|^\tau}{ (1- |\rho_d|^2)^{-j}} \nonumber \\
& \times \sum_{t=0}^{j-1}
\binom{j-1}{t} \left(\frac{ |\rho_d|^2}{ 1 - |\rho_d|^2}\right)^t
\frac{ \Gamma(u)}{(\tau+t)!}
g_1(u)
\end{align}
and
\begin{align} \label{eq:cijFinal}
c(i,j) &= \frac{ \log_2(e) e^{N_t/\gamma} \Gamma(\tau + i) \,
|\rho_d|^\tau}{ (1- |\rho_d|^2)^{-i}} \nonumber \\
& \times \sum_{t=0}^{i-1}
\binom{i-1}{t} \left(\frac{ |\rho_d|^2}{ 1 - |\rho_d|^2}\right)^t
\frac{ \Gamma(v)}{(\tau+t)!}
g_1(v)
\end{align}
respectively. We evaluate the remaining integral, $a(i,j)$, by using the power
series expansion
\begin{align} \label{eq:besselExp}
I_{\tau} (x) = \sum_{k=0}^\infty \left( \frac{x}{2}
\right)^{\tau+2k} \frac{ 1}{ k! \, (\tau+k)! }
\end{align}
and integrating term by term using (\ref{eq:intIdentExpInt}), to
obtain
\begin{align} \label{eq:JFinal}
a(i,j) &= (\log_2(e))^2 e^{\frac{2 N_t}{\gamma (1-|\rho_d|^2)}}
|\rho_d|^\tau \left( 1 - |\rho_d|^2 \right)^{\tau + i + j}
\nonumber \\
& \times \sum_{t=0}^\infty \frac{ |\rho_d|^{2 t} \, \Gamma(u) \Gamma(v)}{t!
\, (\tau+t)!} g_2(u) g_2(v) \; .
\end{align}
Substituting (\ref{eq:bijFinal}), (\ref{eq:cijFinal}), and
(\ref{eq:JFinal}) into (\ref{eq:Dunline}), we perform some basic
algebraic manipulations to write (\ref{eq:ACF_capacity_start}) as
follows
%
\begin{align} \label{eq:finalACF}
E \left[ \mathcal{I}_0 \mathcal{I}_{d} \right] &=  \frac{(\log_2
(e))^2 e^{2 N_t / \gamma}}{\Gamma_m(n) \Gamma_m(m)}
 \sum_{ r = 1 }^m \sum_{ s = 1 }^m
\det \left( \mathbf{C}_{r, s} (\rho_d) \right)
\end{align}
for $0 < |\rho_{d}| < 1$.  The proof is completed by substituting (\ref{eq:finalACF}) and
(\ref{eq:Cd_Extreme}) into (\ref{eq:ACFToneDiff}), and then
substituting (\ref{eq:ACFToneDiff}), (\ref{eq:SecMomFlatFading}) and
(\ref{eq:MeanFlatFading}) into (\ref{eq:VarEqn}) and simplifying.

\hfill \interlinepenalty500 $\Box$


\section{Joint P.d.f.\ of Arbitrarily Selected Eigenvalues of Subcarrier
Matrices} \label{sec:App1_ArbEv}

\begin{lemma} \label{lem:ArbEV}
Let $\lambda$ and $\omega$ be arbitrarily selected non-zero
eigenvalues of the subcarrier channel matrices $\mathbf{H}_{k}
\mathbf{H}_{k}^\dagger$ and $\mathbf{H}_\ell
\mathbf{H}_\ell^\dagger$ respectively. Then the joint p.d.f.\ of
$\lambda$ and $\omega$ is given by
\begin{align} \label{eq:arbUnorderedPdf}
f (\lambda, \omega) &= \frac{  |\rho_{d}|^{-m (n-1)}}{ \Gamma_m(n)
\Gamma_m(m) m^2 \left(1 - |\rho_{d}|^2 \right)^m } \nonumber \\
& \hspace*{0.5cm} \times \sum_{ r = 1 }^m
\sum_{ s = 1 }^m \det \left( \mathbf{D}_{r, s} \left( \lambda,
\omega \right) \right)
\end{align}
where $d = k - \ell$, $\tau = n-m$, and $\mathbf{D}_{r, s} \left(
\lambda, \omega \right)$ is an $m \times m$ matrix with $(i,j)^{\rm
th}$ element given by (\ref{eq:DUnlineCF}) at the top of the page, where $I_\tau (\cdot)$ is the modified Bessel function of the first
kind \cite[eq.\ (9.6.10)]{abramowitz65}.
\addtocounter{equation}{1}
\end{lemma}
\vspace*{1cm}

\emph{Proof:}
From (\ref{eq:freqCorrCoef}), we see that $\mathbf{H}_k
\mathbf{H}_k^\dagger$ and $\mathbf{H}_\ell \mathbf{H}_\ell^\dagger$
are (frequency) \emph{correlated Wishart matrices}.  In
\cite{smith06}, the joint ordered eigenvalue density for matrices of
this general form was evaluated for cases where the correlation
coefficient was real. Extending this result to complex correlation
coefficients, and to unordered eigenvalues, we obtain the joint
eigenvalue density
\begin{align} \label{eq:unorderedJointpdf}
f_u \left(\mathbf{\underline{\lambda}},\mathbf{\underline{\omega}}
\right) &= \frac{1}{m!^2} \frac{ |\rho_{d}|^{-m (n-1)}}{\Gamma_m(n)
\Gamma_m(m) \left(1 - |\rho_{d}|^2 \right)^m } \nonumber \\
& \hspace*{0.2cm} \times \exp \left( - \frac{
\sum_{t=1}^m \left( \lambda_t + \omega_t \right) }{1-|\rho_{d}|^2}
\right) \Delta_m \left(\mathbf{\underline{\lambda}}\right) \Delta_m
\left(
\mathbf{\underline{\omega}} \right) \nonumber \\
& \hspace*{0.2cm} \times \det \left( ( \lambda_i \omega_j )^{
\frac{\tau}{2}} I_{\tau} \left( \frac{ 2 |\rho_{d}|}{1-
|\rho_{d}|^2} \sqrt{\lambda_i \omega_j } \right) \right)
\end{align}
where $\Delta_m(\cdot)$ is a Vandermonde determinant, defined as
\begin{align}
\Delta_m \left(\mathbf{\underline{\lambda}}\right) = \prod_{i<j}^m
\left( \lambda_j - \lambda_i \right) = \det \left( \lambda_i^{j-1}
\right) \; .
\end{align}
Note that the extension from ordered to unordered eigenvalues simply
involved the addition of the leading $1/m!^2$ factor in
(\ref{eq:unorderedJointpdf}), whereas the extension from real to
complex correlation coefficients is trivial, and the proof is
omitted.

To evaluate (\ref{eq:arbUnorderedPdf}) we marginalize
(\ref{eq:unorderedJointpdf}) as follows
\begin{align} \label{eq:jointDensity}
& f (\lambda, \omega) \nonumber \\
& \hspace*{0.05cm} = \int_{\lambda_2} \cdots \int_{\lambda_m}
\int_{\omega_2} \cdots \int_{\omega_m} f_u
\left(\mathbf{\underline{\lambda}},\mathbf{\underline{\omega}}
\right) {\rm d}\lambda_2 \cdots {\rm d}\lambda_m {\rm d}\omega_2
\cdots {\rm d}\omega_m
\end{align}
where we have let $\lambda_1 = \lambda$ and $\omega_1 = \omega$. We
evaluate these integrals by first expanding the Vandermonde
determinants in (\ref{eq:unorderedJointpdf}) according to
\begin{align} \label{eq:VandDetExpand}
& \Delta_m \left(\mathbf{\underline{\lambda}}\right) \Delta_m \left(
\mathbf{\underline{\omega}} \right) = \sum_{ \alpha } (-1)^{{\rm
per}(\alpha)} \prod_{i=1}^m \lambda_i^{\alpha_i - 1} \nonumber \\
& \hspace*{3cm} \times \sum_{ \beta }
(-1)^{{\rm per}(\beta)} \prod_{j=1}^m \omega_j^{\beta_j - 1}
\end{align}
where the sums are over all permutations $\alpha = \left\{\alpha_1,
\ldots, \alpha_m \right\}$ and $\beta = \left\{\beta_1, \ldots,
\beta_m \right\}$ of $\left\{1, \ldots, m \right\}$, and $(-1)^{{\rm
per}(\alpha)}$ and $(-1)^{{\rm per}(\beta)}$ denote the signs of the
permutations. Substituting (\ref{eq:VandDetExpand}) and
(\ref{eq:unorderedJointpdf}) into (\ref{eq:jointDensity}) yields
\begin{align}
& f (\lambda, \omega) \nonumber \\
& \hspace*{0.2cm} = \int_{\lambda_2} \cdots  \int_{\omega_m}
\frac{  |\rho_{d}|^{-m (n-1)}}{\Gamma_m(n) \Gamma_m(m) (m!)^2
\left(1 - |\rho_{d}|^2 \right)^m } \nonumber \\
& \hspace*{0.8cm} \times \exp \left( - \frac{ \sum_{t=1}^m
\left( \lambda_t + \omega_t \right) }{1-|\rho_{d}|^2} \right) \sum_{
\alpha } (-1)^{{\rm per}(\alpha)} \nonumber \\
& \hspace*{0.8cm} \times \prod_{i=1}^m \lambda_i^{\alpha_i - 1}
\sum_{ \beta } (-1)^{{\rm per}(\beta)} \prod_{j=1}^m
\omega_j^{\beta_j - 1} \nonumber \\
& \hspace*{0.8cm} \times \det \left( ( \lambda_i \omega_j )^{
\frac{\tau}{2}} I_{\tau} \left( \frac{ 2 |\rho_{d}|}{1-
|\rho_{d}|^2} \sqrt{\lambda_i \omega_j } \right) \right) {\rm
d}\lambda_2  \cdots
{\rm d}\omega_m \nonumber \\
& \hspace*{0.2cm} = \frac{  |\rho_{d}|^{-m (n-1)}}{\Gamma_m(n) \Gamma_m(m) (m!)^2
\left(1 - |\rho_{d}|^2 \right)^m } \nonumber \\
& \hspace*{0.8cm} \times \sum_{ \alpha } \sum_{ \beta }
(-1)^{{\rm per}(\alpha) + {\rm per}(\beta)} \nonumber \\
& \hspace*{0.8cm} \times \int_{\lambda_2} \cdots  \int_{\omega_m} \det
\left( a(\lambda_i, \omega_j, \alpha_i, \beta_j) \right) {\rm
d}\lambda_2 \cdots {\rm d}\omega_m
\end{align}
where
\begin{align} \label{eq:aDefn}
a(\lambda_i, \omega_j, \alpha_i, \beta_j) &=
\lambda_i^{\frac{\tau}{2}+\alpha_i-1}
\omega_j^{\frac{\tau}{2}+\beta_j-1} e^{-\frac{
\lambda_i}{1-|\rho_{d}|^2} } \nonumber \\
& \times e^{ -\frac{ \omega_j}{1-|\rho_{d}|^2} }
I_{\tau} \left( \frac{ 2 |\rho_{d}| }{1-|\rho_{d}|^2} \sqrt{
\lambda_i \omega_j } \right) \; .
\end{align}
Expanding the determinants, integrating term by term, and re-forming
determinants, we obtain
\begin{align} \label{eq:flamommid}
f (\lambda, \omega) & = \frac{  |\rho_{d}|^{-m (n-1)}}{\Gamma_m(n)
\Gamma_m(m) (m!)^2 \left(1 - |\rho_{d}|^2 \right)^m } \nonumber \\
& \times \sum_{ \alpha
} \sum_{ \beta } (-1)^{{\rm per}(\alpha) + {\rm per}(\beta)} \det
\left( \tilde{\mathbf{D}}_{\alpha,\beta} \left( \lambda, \omega
\right) \right)
\end{align}
where $\tilde{\mathbf{D}}_{\alpha,\beta} \left( \cdot \right)$ is an
$m \times m$ matrix with $(i,j)$th element given by (\ref{eq:DtildeCF}) at the top of the page.
\addtocounter{equation}{1}
Reordering rows and columns yields
\begin{align} \label{eq:Dalphabeta}
\det \left( \tilde{\mathbf{D}}_{\alpha,\beta} \left( \lambda, \omega
\right) \right) = (-1)^{{\rm per}(\alpha) + {\rm per}(\beta)} \det
\left( \mathbf{D}_{\alpha_1,\beta_1} \left( \lambda, \omega \right)
\right)
\end{align}
where
\begin{align} \label{eq:DetijAlphaBeta1}
\left(\mathbf{D}_{\alpha_1,\beta_1} \left( \lambda, \omega \right)
\right)_{i,j} = \left\{
\begin{array}{ll}
 a(\lambda, \omega, i, j) & \text{for} \; \; i = \alpha_1, \; j = \beta_1 \\
 b(\lambda, i, j) & \text{for} \; \;  i = \alpha_1, \; j \neq \beta_1 \\
 c(\omega, i, j) & \text{for} \; \;  i \neq \alpha_1, \; j = \beta_1 \\
 d(i, j) & \text{for} \; \; i \neq \alpha_1, \; j \neq \beta_1 \\
\end{array}
\right. \; .
\end{align}
Applying (\ref{eq:Dalphabeta}) in (\ref{eq:flamommid}) we can
further simplify as follows
\begin{align}
f (\lambda, \omega) &= \frac{  |\rho_{d}|^{-m (n-1)}}{ \Gamma_m(n)
\Gamma_m(m) (m!)^2 \left(1 - |\rho_{d}|^2 \right)^m } \nonumber \\
& \hspace*{2cm} \times \sum_{ \alpha
} \sum_{ \beta } \det \left( \mathbf{D}_{\alpha_1,\beta_1} \left(
\lambda, \omega
\right) \right) \nonumber \\
&= \frac{  |\rho_{d}|^{-m (n-1)} \left((m-1)!\right)^2}{\Gamma_m(n)
\Gamma_m(m) (m!)^2 \left(1 - |\rho_{d}|^2 \right)^m } \nonumber \\
& \hspace*{2cm} \times \sum_{
\alpha_1 = 1 }^m \sum_{ \beta_1 = 1 }^m \det \left(
\mathbf{D}_{\alpha_1,\beta_1} \left(
\lambda, \omega \right) \right) \nonumber \\
&= \frac{  |\rho_{d}|^{-m (n-1)}}{\Gamma_m(n) \Gamma_m(m) m^2
\left(1 - |\rho_{d}|^2 \right)^m } \nonumber \\
& \hspace*{2cm} \times \sum_{ r = 1 }^m \sum_{ s = 1 }^m
\det \left( \mathbf{D}_{r, s} \left( \lambda, \omega \right) \right)
\; .  \label{eq:flamom}
\end{align}

The result now follows by combining (\ref{eq:aDefn}),
(\ref{eq:DetijAlphaBeta1}) and (\ref{eq:flamom}), and by evaluating
the integrals $b(\cdot)$, $c(\cdot)$ and $d(\cdot)$ inside the
remaining determinant, using the identities \cite{smith06}
\begin{align}
& \int_0^\infty x^{a + \frac{t}{2}-1} e^{-cx} I_t (2 \sqrt{f x}) {\rm
d}x \nonumber \\
& \hspace*{1cm} = \frac{ (t + a - 1)!}{c^{\frac{t}{2} + a}} \left( \frac{f}{c}
\right)^{\frac{t}{2}} e^{\frac{f}{c}} \sum_{r=0}^{a-1}
\binom{a-1}{r} \frac{ \left( \frac{f}{c} \right)^r}{ (t + r)!}
\end{align}
for integers $a$ and $t$, and \cite{gradshteyn65}
\begin{align}
\int_0^\infty x^t e^{-a x} {\rm d}x = \Gamma(t+1) a^{-(t+1)}
\end{align}
for integer $t \geq 0$.

\hfill \interlinepenalty500 $\Box$

\section{Proof of Theorem \ref{th:VarHighSNR}}
\label{app:ProofHighSNRVar}

\emph{Proof:} We start by noting that at high SNR, (\ref{eq:logDet})
approaches
\begin{align} \label{eq:HighSNRExpansion}
\mathcal{I}_k &= \log_2 \det \left( \frac{\gamma}{N_t} \mathbf{W}_k
\right)
\end{align}
where is an $m \times m$ complex Wishart matrix given by
\begin{align}
\mathbf{W}_k = \left\{
\begin{array}{ll}
 \mathbf{H}_k \mathbf{H}_k^\dagger & \text{for} \; \; N_r \le N_t \\
\mathbf{H}_k^\dagger \mathbf{H}_k  & \text{for} \; \; N_r > N_t
\end{array}
\right. \; .
\end{align}

Substituting (\ref{eq:HighSNRExpansion}) into (\ref{eq:VarEqn}) and
using (\ref{eq:ACFToneDiff}), we write the variance of the MIMO-OFDM
mutual information at high SNR as follows
\begin{align}
{\rm Var}^\infty(\mathcal{I}_{\rm ofdm}) &= \biggl( \frac{2}{N^2}
\sum_{d=1}^{N-1} (N - d) \nonumber \\
& \times E \left[ \log_2 \det \left(
\frac{\gamma}{N_t} \mathbf{W}_0  \right) \log_2 \det \left(
\frac{\gamma}{N_t} \mathbf{W}_{d}  \right) \right] \nonumber
\\
& + \frac{1}{N} E \left[ \left( \log_2 \det \left(
\frac{\gamma}{N_t} \mathbf{W}_0 \right) \right)^2 \right] \nonumber \\
& - E^2
\left[ \log_2 \det \left( \frac{\gamma}{N_t} \mathbf{W}_0  \right)
\right] \biggr) \; . \label{eq:VarHighSNRExp}
\end{align}
Noting that
\begin{align}
\log_2 \det \left( \frac{\gamma}{N_t} \mathbf{W}_0  \right) =
m\log_2 \left(\frac{\gamma}{N_t}\right) + \log_2 \det \left(
\mathbf{W}_0 \right)
\end{align}
we apply some simple algebra to (\ref{eq:VarHighSNRExp}) and find
that the terms involving $\gamma$ cancel perfectly, leaving
\begin{align}
{\rm Var}^\infty(\mathcal{I}_{\rm ofdm}) &= \biggl( \frac{2}{N^2}
\sum_{d=1}^{N-1} (N - d) \nonumber \\
& \times E \left[ \log_2 \det \left( \mathbf{W}_0
\right) \log_2 \det \left( \mathbf{W}_{d} \right) \right] \nonumber
\\
& + \frac{1}{N} E \left[ \left( \log_2 \det \left(
\mathbf{W}_0 \right) \right)^2 \right] \nonumber \\
& - E^2 \left[ \log_2 \det
\left( \mathbf{W}_0  \right) \right] \biggr)
\label{eq:VarHighSNRExp_Cancelled}
\end{align}
Since $\mathbf{W}_0$ is a complex Wishart matrix, we invoke results
from \cite{grant02} to give
\begin{align}
&E \left[  \log_2 \det \left(
\mathbf{W}_0 \right) \right] = \log_2 (e) \sum_{t=0}^{m-1} \psi (n - t) \label{eq:highSNRWishartMean} \\
& E \left[  \left( \log_2 \det \left( \mathbf{W}_0 \right) \right)^2
\right]  ) \nonumber \\
& \hspace*{0.5cm} = ( \log_2 (e) )^2 \left( \sum_{t=0}^{m-1} \psi'(n-t) +
\sum_{t=0}^{m-1} \psi (n - t) \right)\; .
\label{eq:highSNRWishartVariance}
\end{align}
We now consider the remaining expectation $E \left[ \log_2 \det
\left( \mathbf{W}_0 \right) \log_2 \det \left( \mathbf{W}_{d}
\right) \right]$ in (\ref{eq:VarHighSNRExp_Cancelled}). For the
extreme cases of $\rho_d = 0$ and $\rho_d = 1$, this is directly
obtained from (\ref{eq:highSNRWishartMean}) and
(\ref{eq:highSNRWishartVariance}) respectively.  The main challenge
is to obtain a closed-form finite sum expression for $0 < |\rho_d| <
1$.

\begin{figure*}[!t]
\setcounter{mytempeqncnt}{\value{equation}}
\setcounter{equation}{96}
\begin{align} \label{eq:Ckell_defn_highSNR_inf_sum}
\left(\mathbf{\tilde{C}}_{r, s}(\rho_d) \right)_{i,j} = \left\{
\begin{array}{ll}
\eta_{i,j} (1, \rho_d)  & \text{for} \; \; i \neq r, \; j \neq s \\
 \eta_{i,j} \left( \psi(z), \rho_d \right)  & \text{for} \; \;  i = r, \; j \neq s \\
|\rho_d|^{2(i-j)} \,   \eta_{j,i} \left( \psi(z), \rho_d \right) & \text{for} \; \;  i \neq r, \; j = s \\
\frac{(1-|\rho_d|^2)^{z}}{|\rho_d|^{2 (j-1)}} \sum_{t=0}^\infty
\frac{ |\rho_d|^{2 t} \Gamma(u) \Gamma(v) ( h(\rho_d) + H(u-1) )\, (
h(\rho_d) + H(v-1) ) }{t! (\tau + t)!}  & \text{for} \; \; i = r, \;
j = s
\end{array}
\right.
\end{align}
\setcounter{equation}{\value{mytempeqncnt}}
\hrulefill
\vspace*{4pt}
\end{figure*}

We start by following the same procedure as used in
(\ref{eq:capacityCrossCorr})-(\ref{eq:Dunline}) in the proof of
Theorem \ref{th:exactVarAllSNR}, which yields
\begin{align} \label{eq:ACF_capacity_start2}
& E \left[ \log_2 \det \left( \mathbf{W}_0 \right) \log_2 \det
\left(
\mathbf{W}_{d} \right) \right] \nonumber \\
 & \hspace*{0.5cm}= \frac{
|\rho_d|^{-m (n-1)}}{\Gamma_m(n) \Gamma_m(m) \left(1 - |\rho_d|^2
\right)^m } \sum_{ r = 1 }^m \sum_{ s = 1 }^m \det \left(
\mathbf{\bar{D}}_{r, s} \right) \;
\end{align}
for $0 < |\rho_{d}| < 1$, where $\mathbf{\bar{D}}_{r, s}$ is an $m \times m$ matrix with
entries corresponding to (\ref{eq:Dunline}), but with the
$\alpha(\cdot)$ functions replaced with
\begin{align}
\tilde{\alpha}(x) = \log_2 (x) \; .
\end{align}
We now evaluate the integrals for the elements of
$\mathbf{\bar{D}}_{r, s}$ corresponding to $b(i,j)$ and $c(i,j)$ in
(\ref{eq:Dunline}), using the identity \cite[eq.
(4.352.1)]{gradshteyn65}
\begin{align} \label{eq:intIdentPsi}
& \int_0^\infty x^{q-1} e^{-bx} \ln (x) {\rm d}x \nonumber \\
&\hspace*{1cm} = \frac{\Gamma(q)}{b^q} \left( \psi(q) - \ln(b) \right) \; . \; \;
\hspace*{0.2cm} q
> 0, \; \; b > 0
\end{align}
This gives
\begin{align} \label{eq:bijFinalHighSNR}
b(i,j) &= \frac{ \log_2(e) \Gamma(\tau + j) \, |\rho_d|^\tau}{ (1-
|\rho_d|^2)^{-j}} \nonumber \\
& \times \sum_{t=0}^{j-1} \binom{j-1}{t} \left(\frac{
|\rho_d|^2}{ 1 - |\rho_d|^2}\right)^t \frac{ \Gamma(u)
\psi(u)}{(\tau+t)!}
\end{align}
and
\begin{align} \label{eq:cijFinalHighSNR}
c(i,j) &= \frac{ \log_2(e) \Gamma(\tau + i) \, |\rho_d|^\tau}{ (1-
|\rho_d|^2)^{-i}} \nonumber \\
& \times \sum_{t=0}^{i-1} \binom{i-1}{t} \left(\frac{
|\rho_d|^2}{ 1 - |\rho_d|^2}\right)^t \frac{ \Gamma(v)
\psi(v)}{(\tau+t)!} \; .
\end{align}
To evaluate the remaining integrals in $\mathbf{\bar{D}}_{r, s}$,
i.e.\ for the elements $a(i,j)$, we use (\ref{eq:besselExp}) and
(\ref{eq:intIdentPsi}) to obtain
\begin{align} \label{eq:aijFinalHighSNR}
a(i,j) &= (\log_2(e))^2 |\rho_d|^\tau \left( 1 - |\rho_d|^2
\right)^{\tau + i + j} \\
& \hspace*{0cm} \times \sum_{t=0}^\infty \frac{ |\rho_d|^{2 t} \,
\Gamma(u) \Gamma(v)}{t! \, (\tau+t)!} \left( H(u-1) + h(\rho_d)
\right) \nonumber \\
& \times \left( H(v-1) + h(\rho_d) \right) \; .
\end{align}
Next we use (\ref{eq:bijFinalHighSNR})-(\ref{eq:aijFinalHighSNR}) in
(\ref{eq:ACF_capacity_start2}), and perform some basic
simplifications to obtain
\begin{align} \label{eq:CrossCorrHighSNR}
& E \left[ \log_2 \det \left( \mathbf{W}_0 \right) \log_2 \det \left(
\mathbf{W}_{d} \right) \right] \nonumber \\
& \hspace*{1cm} = \frac{(\log_2 (e))^2}{\Gamma_m(n)
\Gamma_m(m)}  \sum_{ r = 1 }^m \sum_{ s = 1 }^m \det \left(
\mathbf{\tilde{C}}_{r, s} (\rho_d) \right)
\end{align}
where $\mathbf{\tilde{C}}_{r, s}(\rho_d)$ is an $m \times m$ matrix
with $(i,j)^{\rm th}$ element given by (\ref{eq:Ckell_defn_highSNR_inf_sum}) at the top of the next page.  \addtocounter{equation}{1}
The expression (\ref{eq:VarHighSNRFinal}) follows by using
(\ref{eq:CrossCorrHighSNR}), (\ref{eq:highSNRWishartVariance}), and
(\ref{eq:highSNRWishartMean}) in (\ref{eq:VarHighSNRExp_Cancelled}).

To complete the proof we must express the infinite summation in
(\ref{eq:Ckell_defn_highSNR_inf_sum}) in the simplified finite-sum form of
(\ref{eq:Ctilde_rs_Simple}). This simplification requires
significant algebraic manipulations, which we now detail.
Start by recalling the definitions $u = t + \tau + i$ and $v = t +
\tau + j$, and writing the infinite sum in
(\ref{eq:Ckell_defn_highSNR_inf_sum}) as follows
\begin{align} \label{eq:Dkl_infSum}
\left(\mathbf{\tilde{C}}_{r, s}(\rho_d) \right)_{i,j} =
\frac{(1-|\rho_d|^2)^{z}}{|\rho_d|^{2 (j-1)}}
\mathcal{S}(|\rho_d|^2)
\end{align}
where
\begin{align} \label{eq:S_sumDefnFirst}
\mathcal{S}(x) &\defeq \sum_{t=0}^\infty \frac{ x^t (\tau+t+i-1)!
(\tau+t+i-1)! }{t! (\tau + t)!} \nonumber \\
& \hspace*{1cm} \times ( h(\sqrt{x}) + H(\tau+t+i-1) )\, \nonumber \\
& \hspace*{1cm} \times (
h(\sqrt{x}) + H(\tau+t+j-1) ) \; .
\end{align}
Note that the series (\ref{eq:S_sumDefnFirst}), and those that follow below, are convergent for $|x| < 1$ (a condition which holds in (\ref{eq:Dkl_infSum})).

Now, (\ref{eq:S_sumDefnFirst}) can be written as
\begin{align} \label{eq:S_sumDefn}
\mathcal{S}(x) &= h^2(\sqrt{x}) \mathcal{S}_1(1, 1, x) + h(\sqrt{x})
\biggl( \mathcal{S}_1(H(i), 1, x) \nonumber \\
& + \mathcal{S}_1(1, H(j), x) \biggr)
+ \mathcal{S}_1(H(i), H(j), x)
\end{align}
where
\begin{align}
& \mathcal{S}_1( f_1(i), f_2(j), x) \nonumber \\
& \hspace*{1cm} \defeq \sum_{t=0}^\infty \frac{
x^t (\tau + t + i -1)! (\tau + t + j -1)! }{t! (\tau + t)!} \nonumber \\
& \hspace*{1.5cm} \times f_1(\tau
+ t + i -1) f_2(\tau + t + j -1) \;
\end{align}
for arbitrary functions $f_1$ and $f_2$. We now consider each of the
infinite sums in (\ref{eq:S_sumDefn}) in turn.

First consider $\mathcal{S}_1(1, 1, x)$. Following a similar
general approach to that used in \cite{wang06}, 
we perform the following sequence of operations\footnote{Note that
for this particular case, a finite expression could be also found by
directly matching the infinite series to a hypergeometric function,
and using associated identities; something which cannot be done in
the other cases.}
\begin{align}
\mathcal{S}_1  ( 1, 1, x) &=  \sum_{t=0}^\infty \frac{ x^t
(\tau + t + i - 1)! (\tau + t + j - 1)! }{t! (\tau + t)!}  \nonumber \\
&= \frac{{\rm d}^{\tau+i-1}}{{\rm d}x^{\tau+i-1}} \sum_{t=0}^\infty
\frac{ x^{t+\tau+i-1} (\tau + t + j - 1)! }{(\tau + t)!}  \nonumber \\
&= \frac{{\rm d}^{\tau+i-1}}{{\rm d}x^{\tau+i-1}}
\sum_{t=\tau}^\infty \frac{ x^{t+i-1} (t + j - 1)! }{t!}  \nonumber \\
&= \frac{{\rm d}^{\tau+i-1}}{{\rm d}x^{\tau+i-1}} \sum_{t=0}^\infty
\frac{ x^{t+i-1} (t + j - 1)! }{t!}  \nonumber \\
&= \frac{{\rm d}^{\tau+i-1}}{{\rm d}x^{\tau+i-1}} x^{i-1}
\sum_{t=0}^\infty \frac{ x^{t} (t + j - 1)! }{t!} \nonumber \\
&= \frac{{\rm d}^{\tau+i-1}}{{\rm d}x^{\tau+i-1}} \left( x^{i-1}
\frac{{\rm
d}^{j-1}}{{\rm d}x^{j-1}} \sum_{t=0}^\infty x^{t+j-1} \right)  \nonumber \\
&= \frac{{\rm d}^{\tau+i-1}}{{\rm d}x^{\tau+i-1}} \left( x^{i-1}
\frac{{\rm d}^{j-1}}{{\rm d}x^{j-1}} \sum_{t=0}^\infty x^{t} \right)
\; . \label{eq:SumManip}
\end{align}
Via application of the Leibnitz formula, it can be shown that
\begin{align} \label{eq:S1_General}
& \mathcal{S}_1  ( 1, 1, x) \nonumber \\
& \hspace*{0.2cm} = \sum_{b = 0}^{i-1} \binom{ \tau + i -
1}{\tau + b} \frac{ (i-1)!}{b!} x^{b} \frac{ {\rm d}^{\tau+j+b-1}}{
{\rm d}x^{\tau+j+b-1}} \left( \sum_{t=0}^\infty x^{t} \right)
\nonumber \\
& \hspace*{0.2cm} = \Gamma(\tau + i) \sum_{b = 0}^{i-1} \binom{i - 1}{b} \frac{
x^{b}}{(\tau + b)!} \frac{ {\rm d}^{\tau+j+b-1}}{ {\rm
d}x^{\tau+j+b-1}} \left( \sum_{t=0}^\infty x^{t} \right) \; .
\end{align}
Now noting that
\begin{align} \label{eq:S2_FirstSum_NoDeriv}
\sum_{t=0}^\infty x^t = \frac{1}{1-x}, \hspace*{2cm} |x| < 1
\end{align}
with derivatives
\begin{align} \label{eq:S2_FirstSum}
\frac{ {\rm d}^{r}}{ {\rm d}x^{r}} \left( \sum_{t=0}^\infty x^{t}
\right) = \frac{r!}{(1-x)^{r+1}}
\end{align}
we can write (\ref{eq:S1_General}) as follows
\begin{align} \label{eq:S1_FirstResult}
& \mathcal{S}_1  ( 1, 1, x) \nonumber \\
& \hspace*{0.2cm} = \frac{ \Gamma(\tau + i)}{(1-x)^{\tau+j}
} \sum_{b = 0}^{i-1} \binom{i - 1}{b} \left( \frac{ x}{1-x}
\right)^b \frac{(\tau+j+b-1)!}{(\tau + b)!} \nonumber \\
& \hspace*{0.2cm} = \frac{ \Gamma(\tau + i) x^{i-1}}{(1-x)^{z} } \sum_{b = 0}^{i-1}
\binom{i - 1}{b} \left( \frac{ 1-x}{x}
\right)^b (\tau+i-b)_{j-1} \nonumber \\
& \hspace*{0.2cm} = \frac{x^{i-1} \, \eta_{j,i} (1, \sqrt{x})}{(1-x)^z} \; .
\end{align}

Now consider $\mathcal{S}_1(H(i), 1, x)$.  Following the same
sequence of operations as in (\ref{eq:SumManip}) and
(\ref{eq:S1_General}), we find that
\begin{align}
& \mathcal{S}_1(H(i), 1, x) \nonumber \\
& \hspace*{0.1cm} = \frac{{\rm d}^{\tau+j-1}}{{\rm
d}x^{\tau+j-1}} \left( x^{j-1} \frac{{\rm d}^{i-1}}{{\rm d}x^{i-1}}
\sum_{t=0}^\infty x^{t} H (t) \right) \nonumber \\
& \hspace*{0.1cm} = \Gamma(\tau + j) \sum_{b = 0}^{j-1} \binom{j - 1}{b} \frac{
x^{b}}{(\tau + b)!} \frac{ {\rm d}^{\tau+i+b-1}}{ {\rm
d}x^{\tau+i+b-1}} \left( \sum_{t=0}^\infty x^{t} H (t) \right)  .
\label{eq:S1_Hi_deriv}
\end{align}
Now we use \cite{neumann91}
\begin{align} \label{eq:eq:S2_SecondSum_NoDeriv}
\sum_{t=0}^\infty x^t H(t) = \frac{ - \ln (1-x) }{1-x} , \; \; \; |x| < 1
\end{align}
and the corresponding derivatives
\begin{align} \label{eq:SecondSum}
\frac{ {\rm d}^{r}}{ {\rm d}x^{r}} \left( \sum_{t=0}^\infty x^{t}
H(t) \right) = \frac{r!}{(1-x)^{r+1}} \left( H(r) - \ln (1-x)
\right),
\end{align}
to write (\ref{eq:S1_Hi_deriv}) as follows
\begin{align} \label{eq:S1_SecondResult}
& \mathcal{S}_1 ( H(i), 1, x) \nonumber \\
& \hspace*{0.2cm} = \frac{ \Gamma(\tau +
j)}{(1-x)^{\tau+i} } \sum_{b = 0}^{j-1} \binom{j - 1}{b} \left(
\frac{ x}{1-x}
\right)^b \frac{(\tau+i+b-1)!}{(\tau + b)!} \nonumber \\
& \hspace*{1cm} \times \left( H(\tau+i+b-1) - \ln(1-x)  \right) \nonumber \\
& \hspace*{0.2cm} = \frac{ x^{j-1} \Gamma(\tau + j)}{(1-x)^{z} } \sum_{b = 0}^{j-1}
\binom{j - 1}{b} \left( \frac{ 1-x}{x}
\right)^b (\tau+j-b)_{i-1} \nonumber \\
& \hspace*{1cm} \times  \left( H(z-1-b) - \ln(1-x)  \right) \nonumber \\
& \hspace*{0.2cm} = \frac{ x^{j-1} }{(1-x)^z} \left( \eta_{i,j} (H(z-1), \sqrt{x}) -
\ln(1-x) \eta_{i,j} (1, \sqrt{x}) \right) \; .
\end{align}
Now consider $\mathcal{S}_1(1, H(j), x)$.  Using exactly the same
approach as for $\mathcal{S}_1(H(i), 1, x)$, we obtain
\begin{align} \label{eq:S1_ThirdResult}
& \mathcal{S}_1  ( 1, H(j), x) \nonumber \\
& \hspace*{0.2cm} = \frac{x^{i-1}}{(1-x)^z} \left(
\eta_{j,i} (H(z-1), \sqrt{x}) - \ln(1-x) \eta_{j,i} (1, \sqrt{x})
\right) \; .
\end{align}

Finally consider $\mathcal{S}_1(H(i), H(j), x)$. We follow the same
sequence of operations as in (\ref{eq:SumManip}) and
(\ref{eq:S1_General}).  In this case it is convenient to take the
successive derivatives based on the order of $i$ and $j$. In
particular, with $i' = \max(i,j)$ and $j' = \min(i,j)$, we obtain
\begin{align}
& \mathcal{S}_1(H(i), H(j), x) \nonumber \\
 & \hspace*{0.2cm} = \frac{{\rm d}^{\tau+i'-1}}{{\rm
d}x^{\tau+i'-1}} \left( x^{i'-1} \frac{{\rm d}^{j'-1}}{{\rm
d}x^{j'-1}}
\mathcal{S}_2(x) \right) \nonumber \\
& \hspace*{0.2cm} = \Gamma(\tau + i') \sum_{b = 0}^{i'-1} \binom{i' - 1}{b} \frac{
x^{b}}{(\tau + b)!} \frac{ {\rm d}^{\tau+j'+b-1}}{ {\rm
d}x^{\tau+j'+b-1}} \mathcal{S}_2(x) \label{eq:S1Last_Derivs}
\end{align}
where
\begin{align} \label{eq:S2}
\mathcal{S}_2(x) \defeq \sum_{t=0}^\infty x^t H(t) H(t+i'-j') \; .
\end{align}
In this case, in contrast to the previous summations in
(\ref{eq:S2_FirstSum_NoDeriv}) and
(\ref{eq:eq:S2_SecondSum_NoDeriv}), the infinite summation in
(\ref{eq:S2}) cannot be directly expressed in a finite form. To
evaluate this series in finite form, we start by using
(\ref{eq:HarmDefn}) to write
\begin{align} \label{eq:S1_Simplify}
\mathcal{S}_2(x) &= \sum_{t=1}^\infty x^t H(t)
\left( H(t) + \sum_{q = 1}^{i'-j'} \frac{1}{t + q} \right) \nonumber \\
&= \sum_{t=1}^\infty x^t H(t)^2 + \mathcal{S}_3(x) \nonumber \\
&= \frac{{\rm L_{i_2}}(1-x) + \ln^2(1-x)}{1-x} + \mathcal{S}_3 (x)
\end{align}
where ${\rm L_{i_2}}(\cdot)$ is the \emph{dilogarithm} function
\cite[eq.\ (27.7.1)]{abramowitz65}, and $\mathcal{S}_3(\cdot)$ is
given by
\begin{align}
\mathcal{S}_3(x)  \defeq \sum_{q = 1}^{i'-j'} \sum_{t=1}^\infty
\frac{ x^t H(t)}{t + q} \; .
\end{align}
Note that the last line in (\ref{eq:S1_Simplify}) followed by using
an identity from \cite{neumann91}. We now manipulate $\mathcal{S}_3
(\cdot)$ as follows
\begin{align}
\mathcal{S}_3(x)  &= \sum_{q = 1}^{i'-j'} \frac{1}{x^{q}}
\sum_{t=1}^\infty \frac{
x^{t+q} H(t)}{t + q} \label{eq:S3_Firstline} \\
&= \sum_{q = 1}^{i'-j'} \frac{1}{x^{q}} \sum_{t=1}^\infty \int
x^{t+q-1} H(t) {\rm d}x
\nonumber \\
&= \sum_{q = 1}^{i'-j'} \frac{1}{x^{q}} \int x^{q-1} \left(
\sum_{t=1}^\infty   x^t H(t)
\right) {\rm d}x \nonumber \\
&= \sum_{q = 1}^{i'-j'} - \frac{1}{x^{q}} \int \frac{x^{q-1}
\ln(1-x)}{1-x} {\rm d}x \; .
\end{align}
For $q > 1$, consider
\begin{align}
\frac{x^{q-1}}{1-x} &= - x^{q-2} + \frac{x^{q-2}}{1-x} = \ldots
\nonumber \\
&= \frac{1}{1-x} - \sum_{v=1}^{q-1} x^{v-1}, \hspace*{2cm} q > 1,
\end{align}
so therefore
\begin{align} \label{eq:S3_Int}
\mathcal{S}_3(x) &= \sum_{q = 2}^{i'-j'} \frac{1}{x^{q}}
\sum_{v=1}^{q-1} \int x^{v-1} \ln(1-x) {\rm d}x \nonumber \\
& \hspace*{1cm} - \sum_{q =
1}^{i'-j'} \frac{1}{x^{q}} \int \frac{\ln(1-x)}{1-x} {\rm d}x \; .
\end{align}
Using \cite[Eq.\ 2.729]{gradshteyn65}\footnote{There is a missing
$(-1)$ factor in this reference.}
\begin{align}
& \int y^m \ln (1-y) {\rm d}y = \frac{1}{m+1} \biggl( (y^{m+1} - 1)
\ln(1-y) \nonumber \\
& \hspace*{1cm} - \sum_{k=1}^{m+1} \frac{ y^{m-k+2} }{m-k+2} \biggr) + {\rm
const}
\end{align}
and noting that
\begin{align}
\int \frac{ \ln(1-x)}{1-x} {\rm d}x &= - \int \ln(1-x) \frac{{\rm
d}}{{\rm d} x} \ln(1-x) {\rm d}x \nonumber \\
&= - \frac{\ln^2(1-x)}{2} + {\rm const}
\end{align}
we can now express $\mathcal{S}_3(x)$ in finite form as follows
\begin{align} \label{eq:S3_closedform}
\mathcal{S}_3(x) &= \sum_{q = 2}^{i'-j'} \frac{1}{x^q}
\sum_{v=1}^{q-1} \frac{1}{v} \left( (x^v - 1) \ln(1-x) -
\sum_{t=1}^{v} \frac{ x^{t} }{t} \right) \nonumber \\
& \hspace*{1cm} + \frac{ \ln^2(1-x)}{2}
\sum_{q = 1}^{i'-j'} \frac{1}{x^q} \; .
\end{align}
Note that it can be easily verified, using (\ref{eq:S3_Firstline}),
that the integration constant generated in going from
(\ref{eq:S3_Int}) to (\ref{eq:S3_closedform}) is zero. After much
algebraic manipulation, it can be shown that
(\ref{eq:S3_closedform}) reduces to
\begin{align} \label{eq:S3_Final}
\mathcal{S}_3(x) &= \frac{ \ln(1-x) }{2} \sum_{q = 1}^{i'-j'}
\frac{\ln(1-x) }{x^{q}} \nonumber \\
& \hspace*{0.5cm} + \sum_{q=1}^{i'-j'-1} \biggl( \frac{ \ln(1-x) H({i'-j'-q})}{x^{q}}
\nonumber \\
& \hspace*{0.5cm} - \frac{\ln(1-x) H(q)}{x^{q+1}} \nonumber \\
& \hspace*{0.5cm} - \frac{1}{x^q}
\sum_{r = 1}^{i'-j'-q} \frac{ H({r+q-1}) - H({r-1})}{r} \biggr) \; .
\end{align}
Now substituting (\ref{eq:S3_Final}) into (\ref{eq:S1_Simplify}) we
can express $\mathcal{S}_2(x)$ as the finite sum
\begin{align} \label{eq:S2_Simplify}
\mathcal{S}_2 (x) &= \frac{{\rm L_{i_2}}(1-x) + \ln^2(1-x)}{1-x} \nonumber \\
& \hspace*{0.1cm}  +
\frac{ \ln(1-x) }{2} \sum_{q = 1}^{i'-j'} \frac{\ln(1-x)}{x^{q}} \nonumber \\
& \hspace*{0.1cm}
   + \sum_{q=1}^{i'-j'-1} \biggl( \frac{\ln(1-x) H({i'-j'-q})}{x^{q}} - \frac{\ln(1-x) H(q)}{x^{q+1}} \nonumber \\
 &   -  \frac{1}{x^{q}} \sum_{r =
1}^{i'-j'-q} \frac{ H({r+q-1}) - H({r-1})}{r} \biggr) \; .
\end{align}
The corresponding derivatives can be obtained after tedious algebra
as follows
\begin{align}
\frac{ {\rm d}^{r}}{ {\rm d}x^{r}} \mathcal{S}_2 (x) &=
\frac{r!}{(1-x)^{r+1}} \xi_x (r) \label{eq:finalS2Deriv}
\end{align}
where
\begin{align}
\xi_x & (r) =   {\rm L_{i_2}}(1-x) + \ln^2(1-x) - 2 H(r) \ln(1-x) \nonumber \\
& +
\sum_{b=1}^r \left( \frac{2 H(b-1) -
 f_{1,b-1}(x) }{b}  \right) \nonumber \\
& \hspace*{0cm} + \frac{1}{2} \sum_{q=1}^{\delta} \biggl( {
\ln(1-x)}f_{q,r}(x) - \sum_{b=0}^{r-1} \frac{ f_{q,b}(x)}{r-b}
\biggr) \nonumber \\
& \hspace*{0cm} + \sum_{q=1}^{\delta-1} \biggl( H(\delta-q)
f_{q,r}(x) - H(q) f_{q+1,r}(x) + \mu_{q,r}(x) K(q) \biggr)
 \label{eq:SijDefn}
\end{align}
where $\delta = i' - j'$, and recall that ${\rm L_{i_2}}(\cdot)$ is
the \emph{dilogarithm} function \cite[eq.\ (27.7.1)]{abramowitz65}.
Also, $K(\cdot)$ is a constant given by
\begin{align}
K(q) = \sum_{t=1}^{\delta-q} \frac{ H(t+q-1)-H(t-1)}{t} \; ,
\end{align}
and
\begin{align}
f_{q,r}(x) = \sum_{t=0}^{r-1} \frac{ \mu_{q,t}(x)}{r-t} -
\mu_{q,r}(x) \ln(1-x)
\end{align}
where
\begin{align}
\mu_{q,r}(x) = \binom{q+r-1}{r} \frac{ (x-1)^{r+1}}{x^{r+q}} \; .
\end{align}

Substituting (\ref{eq:finalS2Deriv}) into (\ref{eq:S1Last_Derivs})
we obtain
\begin{align}
\mathcal{S}_1 & (H(i), H(j), x) \nonumber \\
 &= \frac{ \Gamma(\tau +
i')}{(1-x)^{\tau+j'} } \sum_{b = 0}^{i'-1} \binom{i' - 1}{b}
\left(\frac{ x}{1-x} \right)^b \nonumber \\
& \hspace*{1cm} \times \frac{(\tau+j'+b-1)!}{(\tau + b)!}
\xi_x (\tau+j'+b-1) \nonumber \\
&= \frac{ x^{i'-1} \Gamma(\tau + i')}{(1-x)^{z} } \sum_{b =
0}^{i'-1} \binom{i' - 1}{b} \left( \frac{ 1-x}{x}
\right)^b \nonumber \\
& \hspace*{1cm} \times (\tau+i'-b)_{j'-1} \xi_x (z-b-1) \nonumber \\
&= \frac{ x^{i'-1} \eta_{j',i'} (\xi_x(z-1), \sqrt{x})}{(1-x)^z} \;
. \label{eq:S1_FourthResult}
\end{align}

Finally, substituting (\ref{eq:S1_FourthResult}),
(\ref{eq:S1_ThirdResult}), (\ref{eq:S1_SecondResult}) and
(\ref{eq:S1_FirstResult}) into (\ref{eq:S_sumDefn}), and then
combining with (\ref{eq:Dkl_infSum}) and simplifying, we obtain the
desired finite-sum expression in (\ref{eq:Ctilde_rs_Simple}).

\hfill \interlinepenalty500 $\Box$


\section{Proof of Theorem \ref{th:VarLowSNR}}
\label{app:ProofLowSNRVar}

\emph{Proof:}
We start by following \cite{kiessling04_asym,hanlen05_opt} and applying a first-order Taylor approximation to
(\ref{eq:logDet}) near $\gamma = 0$ to give
\begin{align} \label{eq:TaylorFirstOrder}
\mathcal{I}_k \approx \log_2 (e) \frac{\gamma}{N_t}  {\rm tr}\left(
\mathbf{H}_k \mathbf{H}_k^\dagger \right) \; . 
\end{align}
Note that, as also mentioned in \cite{kiessling04_asym,hanlen05_opt}, we emphasize that this result is only accurate for the low SNR regime; in general, requiring that the condition $\| ( \gamma / N_t ) \mathbf{H}_k \mathbf{H}_k^\dagger \| < 1$ is satisfied.

Now, substituting (\ref{eq:TaylorFirstOrder}) into (\ref{eq:VarEqn}) and
using (\ref{eq:ACFToneDiff}), we
write the variance of the MIMO-OFDM mutual information at low SNR as follows
\begin{align}
{\rm Var}^0(\mathcal{I}_{\rm ofdm})
&= (\log_2(e))^2 \left(\frac{\gamma}{N_t} \right)^2 \nonumber \\
& \hspace*{-1cm} \times \biggl(
\frac{2}{N^2} \sum_{d=1}^{N-1} (N - d) E \left[ {\rm tr}\left(
\mathbf{H}_0 \mathbf{H}_0^\dagger \right) {\rm tr}\left(
\mathbf{H}_{d} \mathbf{H}_{d}^\dagger \right) \right] \nonumber
\\
& \hspace*{-1cm} + \frac{1}{N} E \left[ {\rm tr}^2 \left(
\mathbf{H}_{\rm flat} \mathbf{H}_{\rm flat}^\dagger \right) \right]
- E^2 \left[ {\rm tr} \left( \mathbf{H}_{\rm flat} \mathbf{H}_{\rm
flat}^\dagger \right) \right] \biggr) \label{eq:VarLowSNRExp}
\end{align}
where $\mathbf{H}_{\rm flat}$ is a flat-fading i.i.d.\ Rayleigh
fading channel matrix. From \cite{lozano03}, we have the following
results
\begin{align}
& E \left[ {\rm tr}\left( \mathbf{H}_{\rm flat} \mathbf{H}_{\rm
flat}^\dagger \right) \right] = N_r N_t \label{eq:traceIdent} \\
& E \left[ {\rm tr}^2 \left( \mathbf{H}_{\rm flat} \mathbf{H}_{\rm
flat}^\dagger \right) \right] = N_r N_t (1 + N_r N_t) \; .
\label{eq:traceSqIdent}
\end{align}
For the remaining expectation in (\ref{eq:VarLowSNRExp}) we write
\begin{align}
& E \left[ {\rm tr}\left( \mathbf{H}_0 \mathbf{H}_0^\dagger \right)
{\rm tr}\left( \mathbf{H}_{d} \mathbf{H}_{d}^\dagger \right) \right] \nonumber \\
& \hspace*{1cm} = E \left[ \sum_{i=1}^{N_r} \sum_{j=1}^{N_t} | ( \mathbf{H}_{0}
)_{i,j}|^2 \; \sum_{k=1}^{N_r} \sum_{\ell=1}^{N_t} |
( \mathbf{H}_{d} )_{k,\ell}|^2 \right] \nonumber \\
&= \sum_{i=1}^{N_r} \sum_{j=1}^{N_t} E \left[  | ( \mathbf{H}_{0}
)_{i,j}|^2 \; | ( \mathbf{H}_{d} )_{i,j}|^2  \right] + (N_r N_t)^2 -
N_r N_t \label{eq:TraceCorrMiddle}
\end{align}
where the second line followed by noting that $E \left[| (
\mathbf{H}_{0} )_{i,j}|^2 \; | ( \mathbf{H}_{d} )_{k,\ell}|^2
\right] = 1$ for all $(i,j) \neq (k,\ell)$. Now using
(\ref{eq:freqCorrCoef}), it can be easily shown that
\begin{align}
& E \left[  | ( \mathbf{H}_{0} )_{i,j}|^2 \; | ( \mathbf{H}_{d}
)_{i,j}|^2  \right] \nonumber \\
& \hspace*{1cm} = |\rho_d|^2 E \left[  | ( \mathbf{H}_{d}
)_{i,j}|^4 \right] + (1-|\rho_d|^2) \, E \left[ | ( \mathbf{E}
)_{i,j} |^2 \right] \nonumber \\
& \hspace*{1cm} = 1 + |\rho_d|^2 \; . \label{eq:HcrossCorrExp}
\end{align}
Substituting (\ref{eq:HcrossCorrExp}) into
(\ref{eq:TraceCorrMiddle}) we find that
\begin{align}
E \left[ {\rm tr}\left( \mathbf{H}_0 \mathbf{H}_0^\dagger \right)
{\rm tr}\left( \mathbf{H}_{d} \mathbf{H}_{d}^\dagger \right) \right]
&= N_r N_t ( | \rho_d |^2 + N_r N_t ) \; .
\label{eq:traceCrossFinal}
\end{align}
The theorem now follows by substituting (\ref{eq:traceCrossFinal}),
(\ref{eq:traceSqIdent}), and (\ref{eq:traceIdent}) into
(\ref{eq:VarLowSNRExp}) and then performing some basic simplifications.

\hfill \interlinepenalty500 $\Box$


\end{appendices}


\end{document}